# Object Space is Embodied

Shan Xu[1†], Xinran Feng[2†], Yuannan Li[2], & Jia Liu[2*]

[1] Faculty of Psychology, Beijing Normal University, Beijing, China
[2] Department of Psychology & Tsinghua Laboratory of Brain and Intelligence, Tsinghua University, Beijing, China.
[†]Equal contribution
[*]Correspondence to: liujiathu@tsinghua.edu.cn (J. Liu)

## Abstract

The perceived similarity between objects has often been attributed to their physical and conceptual features, such as appearance and animacy, and the theoretical framework of object space is accordingly conceived. Here, we extend this framework by proposing that object space may also be defined by embodied features, specifically action possibilities that objects afford to an agent (i.e., affordance) and their spatial relation with the agent (i.e., situatedness). To test this proposal, we quantified the embodied features with a set of action atoms. We found that embodied features explained the subjective similarity among familiar objects along with the objects' visual features. This observation was further replicated with novel objects. Our study demonstrates that embodied features, which place objects within an ecological context, are essential in constructing object space in the human visual system, emphasizing the importance of incorporating embodiment as a fundamental dimension in our understanding of the visual world.

*Keywords*: Object Space, Object Representation, Subjective Similarity, Embodied Cognition, Affordance, Situatedness

# Introduction

*"To see and not to do is not yet to see."*

A 14th century anonymous Japanese Zen master

We consistently perceive certain objects as being more similar to each other than to others. For example, apples are perceived as more similar to oranges than to bananas. This phenomenon is conceptualized within the framework of object space, a theoretical construct representing the underlying mechanism of object representation[1–4]. In this space, the axes represent critical object properties, dictating representational distances among objects. For instance, the perceived similarity between apples and oranges, compared to apples and bananas, may be attributed to their relative proximity in object space along an axis that defines curvature from spiky to stubby. Various studies have explored this abstract object space, for example, by examining corresponding behavioral judgments or neural substrates of human and non-human visual cortex[3,5–12]. These studies have primarily focused on how object space is organized around the visual features (e.g., object size, curvature) and conceptual attributes (e.g., animacy) of objects. However, as Brunswik[13] posited, sensory information alone, devoid of ecological context, is insufficient for ecologically valid perception due to the disparity between proximal stimuli (e.g., a retinal image of a circle) and distal stimuli (e.g., a 3-D apple). Empirical evidence also suggests limitations in confining object space to purely physical and conceptual dimensions, exemplified by the observed activation overlap between hands and tools[14]. Therefore, it is essential to explore whether object space encompasses axes other than physical and conceptual features.

Recent work by our group has identified agents' body size as a crucial measure of objects in visual perception, suggesting an intrinsic link between embodied features and object perception[15]. In addition, features dictating potential interactions in agent-object dyads have been shown to affect object representation, such as object size and weight[16], elongation[17], and the shape and orientation of objects' handles[18–20]. Another piece of evidence is functional fixedness, where an object's previous use in a specific function inhibits agents from discovering new applications for the object in problem-solving tasks[21]. While these studies have illustrated the influence of ecologically meaningful and interaction-related factors, they have not quantified the impact of such factors on object space.

Theories in embodied cognition propose at least two candidates to serve this purpose. The first ecological factor, termed affordance, pertains to the potential manipulations afforded by an object, dictated by its intrinsic properties such as structure, shape, and functionality[22]. For example, a typical affordance feature is grasping postures, such as a power grasp on an apple and a precision grip on a pencil. The second ecological factor, termed situatedness, involves the transient spatial relation between objects and agents during manipulation, such as objects' orientation and location relative to agents[23]. Unlike affordance, situatedness is extrinsic to objects, highlighting the impact of *in situ* sensorimotor processes during agent-environment interactions[24]. That is, while affordance may leverage pre-learned, object-centered knowledge, situatedness is specific to a given instance of manipulation and has to be processed on-the-fly. Both affordance and situatedness resonate with Gibson's original concept of ecological processing[25], and have been shown to be encoded in grasping actions[26] and to affect behavioral responses[27]. In this study, we investigated whether affordance and situatedness constitute essential axes of object space.

To address this question, we investigated the relationship between subjective similarity among objects and similarity in their manipulations. Specifically, we used a behavioral paradigm of subjective similarity judgments[8,28] to measure the object space of various coffee mugs. Concurrently, we developed a structured set of "action atoms" to describe both the affordance and situatedness of a given object during agent-object interactions, allowing for the conversion of manipulations into a series of quantifiable action atoms. Accordingly, the (dis)similarity in embodied features was measured by calculating the difference between group average ratings on the action atoms. Finally, we assessed the association between these two similarities and found that affordance and situatedness of embodied features independently explained the subjective similarity among both familiar and novel objects, suggesting the significance of embodied features of the agent-object dyads in constituting object space. This finding extends the traditional framework of object space, highlighting its multi-faceted nature that includes both physical and embodied dimensions.

## Results

**Quantification of subjective similarity and embodied features**

To characterize object space, we first focused on coffee mugs, a representative category of everyday manipulable objects. This choice was based on two criteria: the diversity in visual appearances (e.g., sizes, height-width ratios, decorations, materials, and textures) and manipulations (e.g., grasping the handle of the mug with fingers, whole-handed cradling the mug body, or grasping the opening of the mug, Figure 1d), as well as the prevalence in daily life that guarantees familiarity across individuals from diverse backgrounds, ages, and lifestyles. We deliberately excluded cross-category stimuli to prevent the variance in semantic and conceptual knowledge from overshadowing that of others.

A triplet odd-one-out task[8] was utilized to construct the object space of the coffee mugs (Figure 1a). In this task, participants were instructed to select a mug that appeared least similar to the other two in each randomly assembled triplet of mugs. The probability $p(i,j)$ that participants did not choose either $Mug_i$ or $Mug_j$ when paired with various third mugs served as a measure of the subjective similarity between $Mug_i$ and $Mug_j$ (for details, see Methods). Thus, the larger the value of $p(i,j)$, the higher the subjective similarity between $Mug_i$ and $Mug_j$.

The resulting representational similarity matrix (RSM), depicted in Figure 1b, illustrated the subjective similarity among various mugs, organized according to clustering based on object-wise subjective similarities. The analysis revealed a trend where mugs clustered according to perspective and handle orientation: mugs viewed from overhead perspectives tended to segregate from those viewed at level perspectives, and mugs with right-oriented handles from those with left-oriented handles (see the legend of Figure 1b for more details). This pattern underscores the influence of embodied features on subjective similarities. Notably, within each cluster, visual attributes reflecting design, color, size, decoration, and texture (glossy or rough) varied, supporting the notion that some embodied features, rather than visual features, may affect object representation.

To further quantify the embodied features of these coffee mugs, we developed a structured set of action atoms to characterize object manipulations (See Figure 1c and d for examples, see

Supplementary Figure S2 for an overview of the organization of the action atoms, and see Supplementary Table S1 for the entire list). These action atoms leveraged existing understanding and theoretical proposals to describe primarily the affordance and situatedness in object manipulations. The action atoms on affordance were adopted either from a hand-centered manipulation taxonomy based on the contact area and force application established in humans[29,30], or from the description of kinematic synergies[31]. Meanwhile, the action atoms on situatedness described variations in the spatial relation between an object and a manipulating agent or an effector hand, which codes an object's orientation relative to the manipulating hand (posture orientation), the movement directions required for manipulation, and the distance of the object to the agent.

In addition, each action atom can be further operationalized by sub-atoms, which specify critical spatial or motor properties of each atom (see Supplementary Table S2 for the entire list). For instance, *Power grasp (palm opposition) with thumb abducted*, an affordance action atom, is the posture we typically adopt to grasp an apple full-handedly. This atom was defined by (1) Thumb abduction: the thumb being opposed to another finger OR the thumb opposed to the back of another finger; (2) Power-ness of grasping: the inner side of the index finger applied force on the object OR the palm applied force on the object OR the inner side of the middle finger applied force on the object; (3) Virtual finger 1: the inner side of the thumb applied force on the object OR the pad of the thumb applied force on the object, and (4) Virtual finger 2 and others: The inner side of the index finger applied force on the object OR the inner side of the middle finger applied force on the object OR the pad of the index finger applied force on the object OR the pad of the middle finger applied force on the object. Similarly, *Hand movement: leftward*, a situatedness action atom, dictates the direction of hand movement in manipulating the object to be leftward. It was further specified by (1) The hand moved between 7 and 11 o'clock on the coronal plane AND (2) The hand moved between 7 and 11 o'clock on the horizontal plane after contacting the object. Detailed descriptions and definitions of this set of action atoms are available in Methods and Supplementary Table S1.

With this set of action atoms, a specific object manipulation was precisely characterized by a combination of these atoms. For instance, for the highlighted agent (the lined silhouette) in Figure 1c, the mug highlighted in the green frame may present affordance action atoms *Thumb-abducted*

*grasping* and *Power-pad grasping with two virtual fingers*, as well as situatedness action atom *Palm direction: outwards* and *Elbow bending*. This means the agent needs to bend the elbow with the palm facing outwards to reach the mug and shape his hand in such a way that the thumb is abducted and gripped with two virtual fingers to handle the mug. In contrast, for the same agent, the mug highlighted in blue may characterize a partially different set of atoms, including rightwards palm direction, elbow stretching, and power-palm grasping.

After establishing the action atoms, we assessed the embodied features of the coffee mugs. To do so, we asked another group of participants to indicate their manipulations of the mugs, such as object contact and force application of each finger, which were subsequently encoded into the action atoms. For instance, for a given mug, if a participant indicated that the thumb will be opposed to the index finger, the palm and the inner side of the index finger and the pad of the thumb would jointly apply force on the mug, this manipulation would score 1 on the action atom *Power grasp (palm opposition) with thumb abducted* (Figure 1d bottom right). In addition, previous studies have suggested potential variations in manipulations between dominant and non-dominant hands as well as across scenarios of functional use or transportation of objects[32,33]. Accordingly, the embodied features were coded for each hand and scenario, respectively. In a given scenario for a given hand, the embodied features of a given mug were calculated as its group-average scores on each action atom, and a representational dissimilarity matrix (RDM) was thus constructed, where each matrix cell represented the corresponding difference between pairs of coffee mugs. With these RDMs, we next characterized the connection between subjective similarity and embodied features.

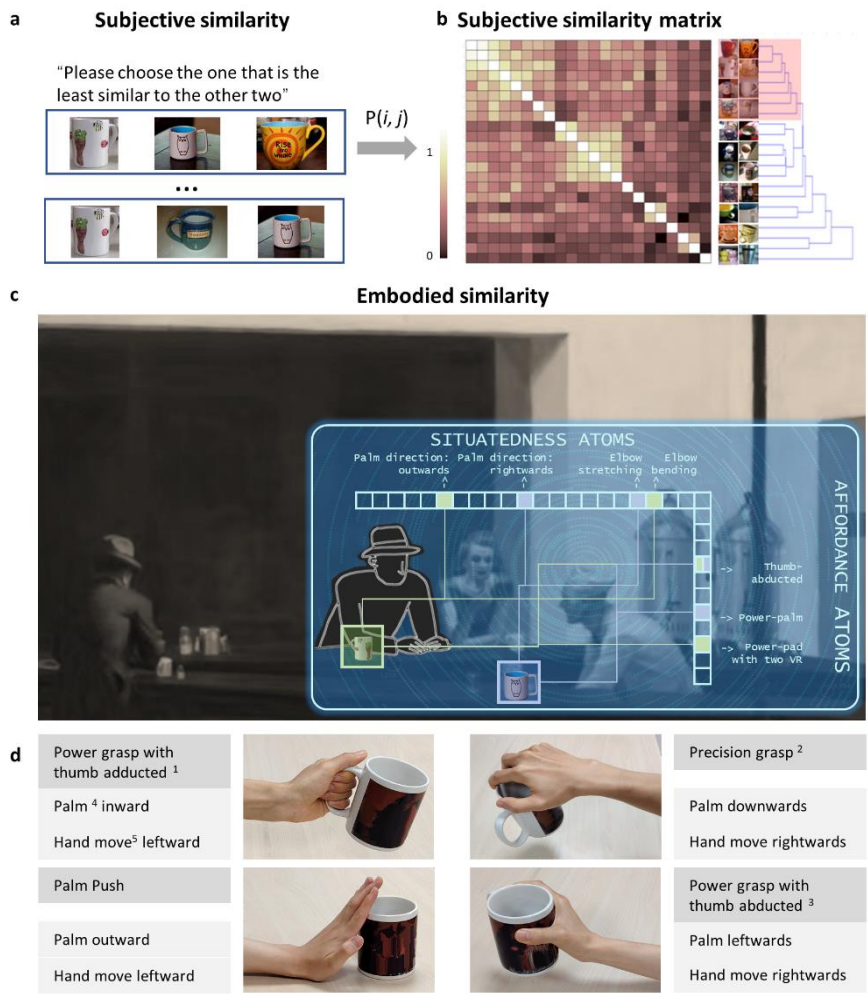

**Figure 1. Quantification of subjective and embodied similarity. a.** Two example trials of the assessment of subjective similarity in a triplet odd-one-out task, where participants identified the mug that appeared least similar to the other two. The probability that two specific mugs appeared in the same trial without being selected as the odd one, $p(i,j)$, indicates the subjective similarity between them. **b.** Subjective similarity matrix of mugs. The $(i,j)$ elements of the matrix correspond to $p(i,j)$. The matrix rows and columns were ordered according to the clustering (the dendrogram) of subjective similarity. The red-shaded mug images showed the composition of an example cluster. Visual inspection suggested that it primarily consists of mugs with handles on the right (with one exception), viewed from a level perspective. **c.** A conceptual illustration of embodied features, adapted from Edward Hopper's *Nighthawk*. For the male figure shown as a lined silhouette, the two mugs afforded different types of grasping, which can be coded into corresponding affordance

(highlighted in the vertical vector on the right) and situatedness (highlighted in the top horizontal vector) atoms. **d**. Example mug manipulation annotated with corresponding action atoms. 1 shortened for *Power grasp (palm opposition) with thumb abducted*. 2 shortened for *Precision grasp (pad opposition) with thumb adducted and the other fingers forming more than two virtual fingers*. 3 shortened for *Power grasp (palm opposition) with thumb abducted.* 4 shortened for *Palm orientation*. 5 shortened for *hand movement direction*.

**Embodied features are parallel to visual features in explaining subjective similarity**

To explore the impact of the embodied features on object representation in the object space, we conducted a linear regression analysis with the vectorized subjective dissimilarity (1 - similarity) matrix as the dependent variable (DV) and the vectorized RDMs of the affordance and situatedness atoms as independent variables (IVs, for details, see Methods). We found that a substantial amount of the variance in the subjective similarity was accounted for by the embodied features (adjusted $R^2$ = 0.57, Figure 2a). To identify the unique contributions of affordance and situatedness, respectively, as well as their shared variance, we decomposed the explained variance in subjective dissimilarity with separate regression analyses. We found the affordance and situatedness IVs were both significant ($p$s < .001), with adjusted $R^2$ values of 0.34 and 0.52, respectively. That is, the affordance and situatedness IVs together explained more than a third of the variance in subjective dissimilarity, and their shared variance attributed to over a fourth of the variance.

Previous studies have shown that visual features influence the agent-object interactions[16–20]. Consequently, the relationship between embodied features and object representation may be attributed to these visual features. In other words, the impact of embodied features on object representation may merely stem from their association with objects' visual features, thereby not contributing novel information to object space. To address this potential confound, we calculated the visual dissimilarity among coffee mugs using a representative deep convolutional neural network (DCNN) model, AlexNet[34,35], designed for object recognition. DCNNs have demonstrated a remarkable resemblance with human and primate ventral visual pathway, in terms of retinotopy, semantic structures, coding scheme, and organization of representational space[3,36–40]. DCNNs provide human-like visual representations free from any impact from the intrinsic embodiment of human perception. We first calculated a dissimilarity matrix for each layer, with each cell representing dissimilarity (1 - the correlation coefficient of activation of all units in a specific layer) between the corresponding pair of coffee mugs. Accordingly, a series of layer-wise activation dissimilarity matrices were constructed to provide a comprehensive description of visual similarity at all visual levels. These matrices served as the visual IVs and were entered into the first level of a hierarchical regression model, with the embodied IVs entered in the second level. This hierarchical regression allowed us to isolate the variance attributable uniquely to the embodied IVs, after controlling for the influence of the visual IVs.

As expected, the visual IVs explained a significant proportion of the variance in subjective dissimilarity among the exemplars within the mug category (adjusted $R^2 = 0.17$, $p < 0.001$, Figure 2b). Importantly, after adjusting for the influence of the visual IVs, the contribution of the embodied IVs to the subjective dissimilarity remained significant ($\Delta R^2 = 0.49$, $p < 0.001$). Further analysis revealed that there was little overlap in the variance accounted for by the visual IVs and by the embodied IVs (Figure 2b). Therefore, this finding dismisses the possibility that the association between object representation and the embodied features was mainly attributable to the visual features. Instead, the embodied features exert a parallel influence on subjective similarity compared to the visual features, suggesting that the embodied features introduce novel dimensions to object representation and contribute to the formation of new axes within the object space.

To explore how the visual and embodied features jointly form object space, we directly compared the representational geometry in the subjective object space derived from subjective similarity judgments on the coffee mugs with that in the object spaces constructed by the visual features exacted from the AlexNet and the embodied features derived from participants' manipulation judgments. Figure 2c presented two-dimensional (2D) visualizations of the representational geometry of the stimuli in the subjective object space (middle), embodied feature space (left), and the visual feature space (right), derived from the non-metric multidimensional scaling (MDS)[41–44] of subjective similarity data and t-distributed stochastic neighbor embedding (t-

SNE) projection of the embodied or visual component scores from corresponding principal component analysis (PCA) analyses. The mug images were framed in the same color scheme as in the dendrogram in Figure 1b, reflecting their clustering in the subjective object space. Visual inspection suggests that the mugs in the first subjective similarity cluster remained close together in the embodied space, and so did their relative segregation from the other two clusters.

To further quantify these observations, we first estimated the dimensionality of the subjective object space by comparing the MDS stress curve (Supplementary Figure S3) with the rule of thumb goodness-of-fit criterion (stress < 0.05)[45]. We found that the coffee mugs were satisfactorily represented in seven dimensions (Supplementary Figure S3a); therefore, in the following analysis, we took the space constructed by seven MDS dimensions as an approximation of the subjective object space, where each object was represented by 7-D coordinates on the seven MDS dimensions. Then, we estimated the dimensionality of the embodied and visual feature spaces based on PCA. The effective dimensionality (ED)[46,47] of the embodied and visual feature spaces were 5.60 and 18.64, respectively (Supplementary Figure S3b), indicating the first 6 embodied components and 19 visual components derived from the PCA could effectively capture variance between objects in the embodied and visual feature spaces, respectively. We therefore limited the following construction of the alternative object spaces within this pool of 25 components.

To measure the similarity between the subjective object space and any alternative space constructed by the combinations of the embodied and visual components, we used Procrustes distances[48] to assess dissimilarity in shape between the manifolds of the mugs embedded in the subjective object space and those embedded in various constructed alternative spaces. Procrustes distance is an established index of similarity of shapes, and has been applied in the comparison of anatomical shapes[49], sensorimotor trajectories[50], and perception and representation geometries[51,52]. Here we intended to compare the manifolds in which objects were represented in a high-dimensional and local Euclidean space[53]. Analyzing the manifold structure reveals the complexity and essential dimensions of object representation[8]. The smaller the Procrustes distances, the more similar the two manifolds and, consequently, the two object spaces. To search for the minimal Procrustes distance, we exhausted all possible combinations from the embodied and visual component pools with a total number equal to or smaller than seven (to either match the dimensionality of the subjective spaces or one of its sub-spaces) (see Methods). As shown in Figure 2d, the Procrustes distances varied among different combinations of the embodied and visual components, with the minimal Procrustes distance achieved with an alternative object space constructed by two embodied components and one visual component to mirror a 3D sub-space of the subjective object space. Consistently, in matching the entire 7-D subjective object space, three embodied components (the first three PCA components) and four visual components collectively generated the best match. Though it is usually difficult to assign semantic labels to PCA components, some of the emerged components seem semantically meaningful. For instance, the first embodied component probably corresponds to the manipulation hand (left or right), with all the mugs with right-pointed handles scoring positive on this component and all but one mug with left-pointed handles scoring negative. The second component likely corresponds to power grasp, with the mug with the highest score on this component being the only one with no handle and therefore, having to be grasped with a power grasp. This finding suggests that the embodied features, often joined by the visual features, are necessary for constructing the object space that represents the subjective similarity among objects. Similar results were replicated with the subjective object space of different dimensionality (Supplementary Figure S4).

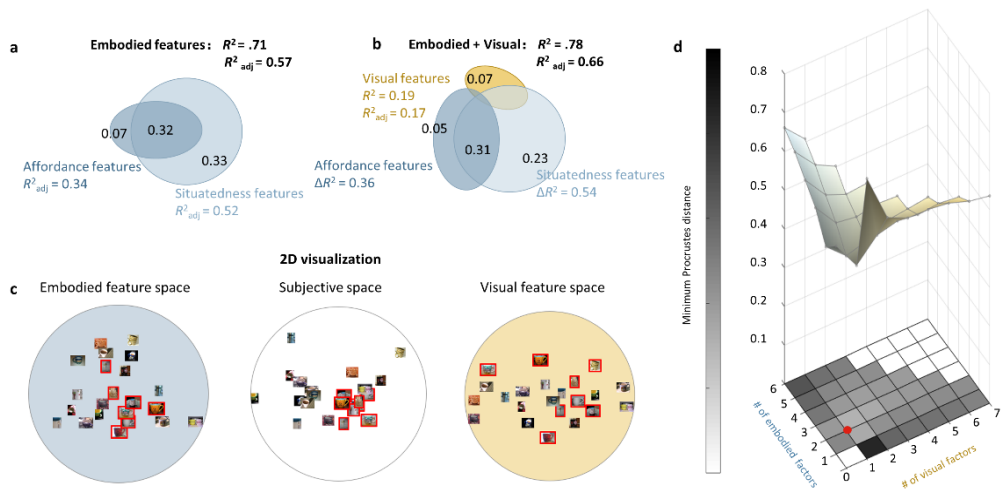

**Figure 2. Embodied features are encoded in subjective similarity. a**. Both the affordance and situatedness features substantially explain the subjective similarity, and around one-third of the variance can be explained by both kinds of features. $R^2_{adj}$ of either category of embodied features

indicates the adjusted $R^2$ when the regression contained solely the dissimilarity vectors of corresponding features as IVs. The numbers in different areas in the Venn diagram indicate the respective proportions of variance of subjective similarity being explained by affordance features only, affordance and situatedness features, and situatedness features only. **b**. Visual similarity also explained a small but significant proportion of variance of the subjective similarity, with little overlap with the embodied IVs. $\Delta R^2$ of either category of embodied features indicates the change of $R^2$ when the dissimilarity vectors of corresponding features were entered into the second layer of the hierarchical regression with the visual IVs in the first layer. The numbers in different areas in the Venn diagram indicate the respective proportions of variance of subjective similarity being explained by affordance features only, affordance and situatedness features, and situatedness features only after taking visual IVs into account. **c**. 2D visualization of the subjective object space (middle), embodied feature space (left), and visual feature space (right), generated from the 2D non-metric multidimensional scaling of subjective similarity data and 2D t-SNE projection of the embodied or visual component scores from corresponding PCA analyses. Data points mark the locations of each mug exemplar in corresponding spaces. Mug images are framed in the same color as in the dendrogram from Figure 1b, to illustrate how their clustering in the subjective object space was preserved or changed in other spaces. **d**. The best match, i.e. the minimum Procrustes distance (z-axis), between the subspaces of the 7D subjective object space and any equal-dimensional alternative space constructed by varied number of embodied (y-axis) and visual components (x-axis). The lowest point on the surface denotes the overall best goodness of fit, which was achieved between a 3D subspace of the subjective object space and an alternative object space constructed jointly by two embodied and one visual component (the red spot on the horizontal plane).

**Representation-manipulation association is independent of experiences**

One may question whether the results we found with coffee mugs are category-specific, influenced by our experiences with this specific object category. To rule out this possibility and further replicate our findings, we examined whether the association between object representation and manipulation remained after removing conceptual and experiential determinants. To do this, we designed a set of novel objects resembling the traditional cube stimuli used in the mental rotation literature[54] but with greater shape complexity (See Supplementary Figure 1 for the entire set). All the novel objects were topologically equivalent to a solid cube, with smooth surface showing a wooden texture (Figure 3a, top). They were presented with identical lighting but with variations in the directions of their long axes and front facets. Additionally, before the collection of subjective similarity data, an example image of a novel object was presented in juxtaposition with an image of an apple as a size reference, and the participants were informed that these novel objects were of a size similar to an apple before both subjective and embodied judgment tasks.

Following the exact procedure used to analyze the coffee mugs, we measured the subjective similarity among these novel objects. Figure 3a illustrates the subjective similarity matrix according to the hierarchical clustering of the similarity vectors of each image. Same with the coffee mugs, the clustering also suggested an impact from manipulation properties. Particularly, we observed that the novel objects consisting of rod-like structures and those of chunky structures tended to segregate, probably due to these two kinds of structures being likely to be grasped with different grasping actions.

To further quantify the embodied features of this set of objects and validate our action atoms of object manipulations, we asked the participants to directly assess the embodied features based on the action atoms (Figure 3b). We found that the association between subjective similarity and embodied features of novel objects replicated the results with coffee mugs. Specifically, regression analysis revealed that the dissimilarity of embodied features of the novel objects explained a significant proportion of the variance in subjective dissimilarity (adjusted $R^2 = 0.42$, Figure 3c), and the magnitude was comparable to that observed with the coffee mugs. Regression restricted to the situatedness or affordance IVs were both significant ($p$s < .001, adjusted $R^2 = 0.21$ and 0.39, respectively), and each explained proportions of variation in subjective similarity similar to that in the coffee-mug analysis.

Furthermore, consistent with the findings from the coffee mugs, the contribution of embodied similarity was dissociable from that of visual similarity for the novel objects (Figure 3d). The visual IVs were significant ($p$s < .001, adjusted $R^2 = 0.07$) but explained a smaller proportion of variation compared to embodied features, presumably because the stimuli varied less in visual features than in manipulation. Again, this pattern is similar to that observed in the analysis of coffee mugs.

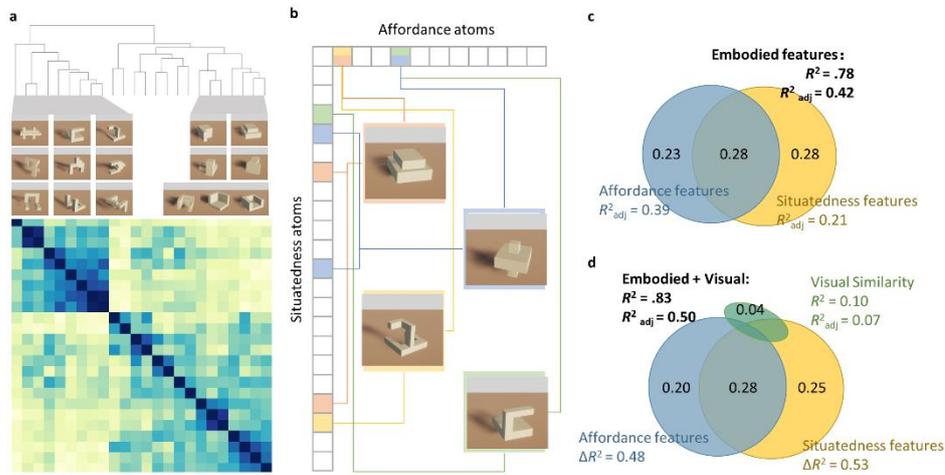

**Figure 3. The experiment on the novel objects. a**. The subjective similarity matrix of novel objects (below). The matrix was arranged in accordance with the hierarchical clustering (above) of the object-wise similarity vectors. The clustering suggested that images were clustered according to object structures that were associated with embodied features. **b**. A conceptual illustration of the application of action atoms on the novel objects. The top two objects share the same affordance atoms but different situatedness atoms. **c**. Substantial variance of subjective similarity between objects can be explained by similarity in embodied features. $R^2_{adj}$ of either category of embodied features indicates the adjusted $R^2$ when the regression contained solely the dissimilarity vectors of corresponding features as IVs. The numbers in different areas in the Venn diagram indicate the respective proportions of variance of subjective similarity being explained by affordance features only, affordance and situatedness features, and situated features only. **d**. Visual similarity also explained a small but significant proportion of variance of subjective similarity. $\Delta R^2$ of either category of embodied features indicate the change of $R^2$ when the dissimilarity vectors of corresponding features were entered into the second layer of the hierarchical regression with the visual IVs in the first layer. The numbers in different areas in the Venn diagram indicates the respective proportions of variance of subjective similarity being explained by affordance features only, affordance and situatedness features, and situatedness features only after taking visual IVs into account. These results replicated that of the coffee-mug analysis reported in Figure 2a and 2b.

## Discussion

The present study examined the idea that object representation is embodied. Together with previous findings[26,32,55–65], our results highlight the impact of object manipulation on object representation. We extended these findings by systematically characterizing the embodied features, which allowed us to quantitatively depict the representation of object manipulation and its role in constructing object space. The regression analysis showed that a significant portion of the variance in subjective similarity was explained by similarity in embodied features, paralleling the similarity in the image space. Furthermore, the manifold analysis revealed that the object space derived from the subjective similarity judgments was best constructed by including axes representing embodied features. Together, these findings suggest a unique contribution of embodied features, parallel to that of image-based visual features, in constructing object space.

With the inclusion of context-sensitive, and particularly, manipulation-related ecological factors, our study enriches the theoretical framework of object space. This enhancement not only helps explain the representational overlap of hands and tools reported in previous studies[14] but also introduces subjectivity in object representation. Thus, object space is not merely defined by object-centered physical and conceptual features but is also embodied and ecologically constrained through agent-world interaction. This is demonstrated by situatedness features derived from agent-object dyads, which made a unique contribution to object similarity[26,27], in addition to object-centered affordance features that are intrinsic and constant in the identity of a given object for a given agent[15,23,56]. The inclusion of situatedness features in representing objects is consistent with theoretical proposals emphasizing the impact of *in situ* sensorimotor processes during the interaction between an agent and its environment[24,25,66].

Notably, our study found that affordance and situatedness were spontaneous rather than knowledge-driven in object representation. When making subjective similarity judgments, the participants were not explicitly asked to consider object manipulation, yet the impact of embodied features still emerged. Additionally, we replicated this finding with a set of novel and meaningless stimuli. Thus, the inclusion of embodied features in object space is not a byproduct of the learned usage of known objects; instead, objects are represented partially according to how they could be manipulated by the viewing agent in the current situation. Accordingly, our study advocates for a dynamic and process-rich perspective on object representation.

This idea underscores the necessity of including the dorsal visual pathway in addition to the ventral temporal cortex for a comprehensive representation of objects, as previous studies on object recognition have primarily focused on the latter. One primary functionality of the dorsal visual pathway is to provide quick, on-the-fly analysis of objects[55,67–70] and guide the online control of object manipulation[71,72], and thus, supplementing the extraction of intrinsic and invariant object features conducted by the ventral visual pathway. Our findings also highlight the limitations of traditional DCNNs in simulating human object recognition, as they rely solely on the visual properties of visual images. Future studies are needed to advance current AI models to incorporate embodied features in representing the world.

One of the major modifications to the theoretical framework of object space introduced by our study is the incorporation of subjectivity, transforming the concept from an object-centered image space to an agent-object dyad representation space. Here we propose a possible mechanism to explain this transformation, based on our findings that the dimensionality of object space was much smaller than that of image space and that relying solely on visual features resulted in a poor goodness of fit with the manifold in subjective object space. Specifically, this transformation from image space to subjective object space is achieved by removing or compressing axes encoding visual features and including axes encoding embodied features (Figure 4). To illustrate this idea, consider four mugs distributed in a 2D image space based on visual features of height and color (Figure 4a). When these four mugs are perceived by an agent, embodied features are automatically incorporated as new axes to construct a higher-dimensional space (Figure 4b). Subsequently, axes encoding visual features that are less relevant to representing objects are removed or compressed to form a lower-dimensional space specific to the agent (Figure 4c). In this example, the orange mugs are pulled apart in object space due to their differences in affordance, while the blue mugs are pulled closer together because of the similarity in the manipulation they afford.

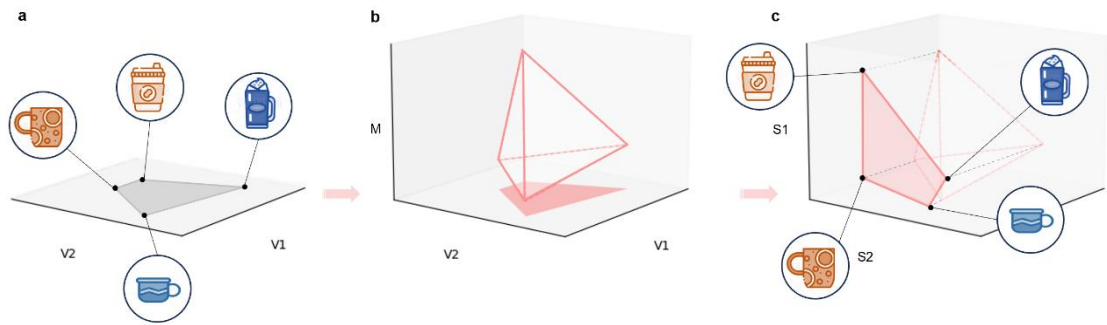

**Figure 4. A conceptual illustration of the speculated evolvement of object space.** In transforming from the objective to the subjective reigns, the original image space (**a**) constructed by visual features (V1 and V2) incorporated embodied features of the agent-object dyads (**b**), and projected to a lower-dimensional but ecologically informed space (**c**, formed by subjective features S1 and S2). In this example, the representation distance between two orange mugs is close in the image space, but it is far in the subjective object space.

This expansion-and-compression transformation of object space results in a representational space that aligns, to some extent, with interaction potentials that the world provides to an agent. This speculated mechanism has critical implications for the functionality of object space. For instance, it suggests that visual representations are generated in a way that facilitates their utility in recognizing and specifying potential interactions between an agent and its environment, even at the cost of objective verisimilitude. Thus, agent-world interactions may serve as a fundamental shaping force of subjective experiences, thereby supporting phenomenological and embodied views of cognition. Future studies on neural manifolds[73,74] are needed to directly examine this speculated mechanism.

One critical methodological advance in this study is the systemization of the taxonomy of object manipulation, which may fuel future studies in both psychology and robotics. To objectively characterize object manipulation and quantify differences in manipulation, our study developed a set of action atoms to describe key aspects of object manipulation. Building upon existing manipulation taxonomies[29–31] and theoretical proposals from diverse fields, including psychology, neuroscience, and robotics, we proposed a relatively structured and systematic description of embodied features. This approach provides a flexible foundation that can be extended to incorporate new understandings of object manipulation taxonomy and tailored to better describe a specific set of objects. For instance, new sets of action atoms can be added for objects with dramatically different sets of affordances or for non-anthropomorphic agents. Importantly, these action atoms, with task-specific fine-tuning, provide a novel tool to transform embodied features of an object into a sequence of action atoms (i.e., action vectors). These action vectors can be processed by AI models built on transformer architectures[75] to generate context-dependent object representations.

In the field of natural language processing, dynamic word representations, also known as contextualized embeddings, allow word representations to be determined in conjunction with specific contexts. Consequently, the embedding of the same word can vary across different sentences and positions, as exemplified by word embeddings in GPTs[76] and other large language models, which overcome the limitations of static word representations (e.g. Word2Vec)[77], such as the polysemy problem[78]. Thus, it is promising that the incorporation of ecological factors into object space, especially embodied features, may lead to similar progress in object recognition and its broader applications. Indeed, incorporating an ecologically valid object space might be a potential approach to building embodied AI, which has been highly anticipated as a potential breakthrough[79] in enabling artificial agents to acquire knowledge by engaging with their immediate environment[80]. Such a human-like subjective perspective of the world may help filter the flood of sensory information into representations ready for generating actions upon the world.

## Materials and Methods

### Participants

Twelve participants (2 males, aged 22-27 years, mean age = 24 years) rated the subjective similarity between coffee mugs, and a non-overlapping set of 19 participants (2 males, aged 22-24 years, mean age = 23 years) rated the manipulation properties of each coffee mug. The participants were undergraduate and graduate students of Beijing Normal University. No restriction was imposed on race, ethnicity, or other socially relevant characteristics. The subjective similarity ratings were pooled across participants, so the sample size of subjective ratings was determined by estimating the required number of ratings to estimate the representational dissimilarity matrix and the number of rating trials the participants could finish within a reasonable amount of time. The size of the manipulation sample was targeted to be between 15 and 20 to get a conventional amount of data points in calculating the across-participants average. Another set of sixteen participants (4 males, aged 19-28 years, mean age = 23 years) rated the subjective similarity and the manipulation properties of novel objects. The participants were undergraduate and graduate students of Tsinghua University. No restriction was imposed on race, ethnicity, or other socially relevant characteristics. All participants were right-handed or mixed-handed neurologically normal volunteers with normal or corrected-to-normal vision.

The study was approved by the Institutional Review Board of Beijing Normal University (BNU, No. 202302220016). The study was performed in accordance with the Declaration of Helsinki. Written informed consent was obtained from all participants before they took part in the experiment, and the participants were compensated financially for their time.

**Materials**

*Everyday objects*

We chose coffee mugs as a representative category of everyday objects due to their variation in appearance, the diversity of applicable manipulations, the prevalence in daily activity, and the absence of skill requirement in manipulation. Twenty-two images were selected from the coffee mug category (n03063599) in the validation dataset of ImageNet Large-Scale Visual Recognition Challenge (ILSVRC) 2012 dataset [81]. ILSVRC images depict objects in a natural background. The images were chosen by the following criteria: (1) 'coffee mug' was among the top five categorizations given by pre-trained AlexNet in pytorch 1.2.0 (https://pytorch.org/, Paszke et al., 2017), (2) it was agreed by two independent raters that a coffee mug was the only main figure without distracting object or person in the image (see Figure 1 for example and see Supplementary Figure S1 for the entire set of stimuli).

*Novel objects*

A set of 24 novel objects was generated to replicate the findings with the coffee mug images (see Figure 3 for example and see Supplementary Figure S1 for the entire set of stimuli). The novel objects were Shepard-and-Metzler style wooden-appearance blocks[54] designed to resemble no known tools and were of a similar size to an apple. They were constructed and captured with the same fixed camera in Unity (2021.3.10f1).

**Procedure**

*Subjective similarity rating*

Subjective similarity was rated using a triplet odd-one-out task similar to that used in a previous study[8].

**Everyday objects.** The sequences of the rating trials were generated and presented using PsychoPy[82]. The images of coffee mugs were presented within a rectangular area with a visual angle of $0.29° \times 0.26°$ on a white background. In each trial, participants viewed three images side by side in a browser window and were instructed to select the one least similar to the other two by pressing the number key 1, 2, or 3 on the keyboard for the left, middle, and right images, respectively. No additional instruction was given. The interval between trials was 300 ms. Each object triplet and the order of triplets were chosen randomly. The participants could terminate the experiment whenever they desired. A total of 6,272 trials were completed, and the participants completed 523 trials on average. The experiment was not further separated into blocks. The whole dataset sampled all the $22 \times 22$ cells in the subjective similarity matrix, with each cell sampled 16 to 27 times across participants. After the experiment, the participants filled out a Chinese version of the Edinburgh Handedness Inventory[83,84].

**Novel objects.** Subjective similarity of novel objects was rated in the same task as that of coffee mugs. The data were collected with Inquisit (https://www.millisecond.com). During practice, participants were first presented with object triplets including either one or two example apples along with randomly selected novel objects. The apples were depicted at the same size as the novel objects with the same lighting and background, serving as a size reference for the novel objects. In the formal experiment, in contrast to the fully randomized sampling of the stimuli pool in each trial of the everyday-object rating, each participant completed 1,012 experimental trials, comprising half of the full combination of object triplets. For each participant, all the trials were pseudorandomly ordered into 11 blocks, with each pair of novel objects being compared 176 times.

*Manipulation judgment*

**Everyday objects.** The participants rated the manipulation properties of each coffee mug image according to a set of 141 questions. The questions were grouped into two parts: one concerned with the manipulation scenarios in which the objects were to be used and the other transported. Within each part, the participants were required to rate the manipulation regarding (1) the part of the hand contacting the objects, (2) the part of the hand applying force during manipulation, (3) the orientation and the posture of hand when contacting the object, and (4) the direction of hand movement involved. In addition, they estimated the distance to the object and provided brief verbal descriptions of the movements and the use of the manipulations respectively. For the questions regarding orientation and movement direction, the participants were required to indicate the direction using the clock dial analogy in the coronal, sagittal, and horizontal planes respectively. For the questions regarding parts of contact or force application, a list of possibilities was presented, and the participants were instructed to make yes or no responses to each one. For postures, questions were asked regarding whether opposition between the thumb and other fingers was involved; if so, the participants were required to select among choices to indicate the finger opposing the thumb, estimate the distance between the opposing fingers and the diameter of the curves formed by the palm, the index finger, and the middle finger, respectively. This set of questions was designed to cover the relatively low-level features of the manipulation, which could be coded into high-level action atoms included in the manipulation alphabet. The manipulation judgment task was self-paced and the participants were required to finish rating all the images within 24 hours. The complete list of the questions can be found in Supplementary Table S1.

Responses to the questions were further encoded into affordance and situatedness atoms (see below). Among the affordance atoms, those corresponding to the contact-and-force-based manipulation were derived by recoding contact and force application responses in the reports (see Supplementary Table S2 for the mapping between raw variables and each manipulation type). Here we coded these atoms from contact and force application reports rather than asking the participants to directly categorize their manipulation to avoid noise from individual differences in action categorization and verb usage. The kinematic-based atom, posture closeness, was encoded based on reported estimation of finger and hand curvatures in handling the object. Among the situatedness atoms, posture orientation and movement direction were derived based on participants' estimation of corresponding orientations and recoded into upwards, downwards, left, right, inwards, and outwards from clockwise direction estimation in the sagittal, coronal, and horizontal planes in the egocentric coordinate system to avoid the impact of noisy direction estimation, and the direction of wrist twist and one regarding elbow flexion were both extracted directly from the response. Perceived distance was also extracted directly from the responses.

**Novel objects**. The participants rated the manipulation properties of each object according to a set of 62 questions. Similar to the coffee mug images, the questions were grouped into manipulation-to-move and manipulation-to-use scenarios, for each hand, respectively. The questions required descriptions regarding (1) the posture of the hand when manipulating the object, (2) the orientation of the hand when manipulating the object, and (3) the direction of hand movement involved. A full list of questions can be seen in Supplementary Table S3.

The manipulation judgment task was presented online via Inquisit, divided into three blocks (with eight novel objects per block). Participants were asked to complete a practice block first to familiarize themselves with the task, in which they rated the example apple presented with the lighting and background as in the following formal blocks. Then, they proceeded to formal blocks, instructed beforehand to consider the novel object as being of the same size as an apple. The participants were required to complete all blocks within 24 hours. The rating was self-paced within each block.

**Action atoms**. The resultant manipulation description scheme consisted of a three-level structure. The first level was a theoretically driven level, based on the theoretical discussion in embodied cognition, in which two broad categories were established: affordance features and situatedness features. According to embodied theories of cognition, object manipulation is collectively dictated by the agent and the objects. An object affords actions suitable to both its physical attributes and the motor capability of the agent. The situated theories of cognition also highlight the importance of the situatedness of object manipulation, i.e. the variable features specific to each instance of object manipulation, such as the agent-object spatial relation. Given a specific agent, the motor capability is fixed, and the mutual relation breaks down into the physical characteristics of the objects and the spatial feature of the objects in relation to the agent. The two categories in the theoretical-driven level corresponded to these two sources of variation in manipulation respectively, and were termed the affordance features (because they are the features frequently studied in the affordance literature) and the situatedness features (because of their *in situ* nature and the emphasis on such features in the situated theories of cognition). The second level was an empirical-driven level, based on the findings from cognitive psychology, cognitive neuroscience, and robotics. This level broke each category (affordance and situatedness) into a set of action atoms. The atoms in the affordance category corresponded to different types of prehensile and nonprehensile object-handling actions in manipulation. In constructing this feature set, we considered well-established taxonomies of prehensile and non-prehensile manipulation actions in the robotics literature[29–31]. We considered both the taxonomies based on the contact points and force application characteristics[29,30] and those based on motor synergies[31], and included 11 features for the everyday objects. Considering the literature on spatial features impacting object manipulation or perception, the situatedness category was broken down into atoms reflecting the movement direction (eight), the distance (one) of the objects in the 3D egocentric coordinates of the agents, and the objects' orientation relative to the manipulating hand (twelve) for the everyday objects. Forming the basis of this level was the operational definitions (sub-atoms coded from the responses in the manipulation judgment task). On top of this three-level structure, the manipulations were parallelly described for both hands and two different object manipulation scenarios (manipulate-to-move vs manipulate-to-use), based on empirical findings suggesting the differentiation between dominant and non-dominant hand and between scenarios in the object manipulation literature[32,33]. The resultant description scheme is shown in Supplementary Figure S2. In principle, this scheme turned a manipulation action into a vector at a desirable level. For instance, to grasp an apple near to the body on the left side and to grasp it with hand palm downwards with a power-palm grasp can be dented as a power-palm grasp, close, left (to the body), down (-ward effector), near manipulation on the second level (See Table 1 for more examples).

It should be noted that we did not argue for the independence or orthogonality of the first two schools of features, and the affordance and situatedness features of a manipulation may be conceivably interdependent, but we proposed that they may cover two key sources of influence on manipulation with respective emphases. We included them to maximize the systematicity and comprehensiveness of our description and comparison in potential application of our scheme to a reasonably wide range of object manipulations.

Table 1. Examples illustrating the mapping between manipulation actions to the affordance features

|   | Affordance feature | Example manipulation with coffee mugs | Other manipulation examples in activities of daily living |
|---|---|---|---|
| a | Power grasp (palm opposition) with thumb abducted | Grasping the body of a mug | Grasping an apple full handedly |
| b | Power grasp (palm opposition) with thumb adducted | Grasping the handle of the mug without the thumb bending towards other fingers | Grasping the railing on the subway with the thumb resting along the railing |
| c | Power grasp (pad opposition) with thumb abducted and the index and (or) middle fingers forming one or two virtual fingers | Grasping the body of an expresso cup with the thumb and the index finger encircling the body of the cup | Grasping a tennis ball with the ball resting on the conjunction between the thumb and the index finger |
| d | Power grasp (pad opposition) with thumb abducted and the other fingers forming more than two virtual fingers | Grasping the saucer under a coffee mug with all the fingers open and all applying force on the saucer | Holding a book with the book cover facing upwards and the thumb pressing on the book cover |
| e | Intermediate side grasp | Holding a spoon to add sugar into a mug with the thumb and the index finger stretching along the spoon and the spoon resting on the index finger | Holding a card with the thumb and the interphalangeal joints of the middle finger with the middle finger loosely stretching |
| f | Precision grasp (pad opposition) with thumb abducted and the index and the middle fingers forming one or two virtual fingers | Picking up an upside-down mug by pinching it with the thumb and the middle finger | Picking up an apple by pinching the stalk with the thumb and the index finger |
| g | Precision grasp (pad opposition) with thumb abducted and the other fingers forming more than two virtual fingers | Grasping the opening of a mug with all the fingers | Picking up a CD with all the fingers open as a cage, holding its lateral surface |
| h | Open hand nonprehensile manipulation with fingers | Pushing a mug away with stretching fingers | Pressing a button with a finger |
| i | Open hand nonprehensile manipulation with palm | Pushing a mug with palm | Hitting a table with the palm |
| j | Open hand nonprehensile manipulation with handback | Pushing a mug with handback | Pushing away sundries on the desk with handback |

For the novel objects, twelve affordance atoms and ten situatedness atoms for each scenario and each hand were extracted from the manipulation rating of novel objects. The number of atoms differed for novel and everyday objects because different numbers of actions were considered applicable to these two sets of objects, and we combined some situatedness atoms, such as hand moving inward and outward, to simplify the analysis for the novel objects.

**Representational similarity analysis**

Following Hebart et al.'s approach[8], for known objects (coffee mugs), a 22 × 22 symmetric matrix was thus generated with the subjective similarity between object $i$ and object $j$ being denoted in cell $(i,j)$, which was defined as the probability of these two objects being chosen to belong together across participants, irrespective of the third object in the triplet. In addition, hierarchical clustering was performed using the *linkage* method in Matlab R2020a (The MathWorks Inc) on the subjective dissimilarity (1 - subjective similarity) of the mugs. The subjective similarity was calculated for the novel objects in the same way.

**Generation of the manipulation similarity matrixes**

For the everyday objects, a manipulation similarity matrix was generated for each action atom. The participants' manipulation judgments for each object were taken as the z-standardized atom scores (see Supplementary Table S2 for the coding scheme of each action atom). Then, for each action atom, a 22 × 22 matrix was generated with the difference in the average score between object $i$ and object $j$ on this feature denoted in cell $(i,j)$, which was the manipulation dissimilarity between this pair of objects on this action atom. Manipulation dissimilarity matrices were generated similarly for the novel objects.

**Generation of visual similarity matrixes**

For both everyday objects and novel objects, the visual similarity of the objects was measured by the Euclidean distance of neural activation of each object pair in AlexNet[34,85]. AlexNet is a feed-forward hierarchical convolutional neural network consisting of five convolutional layers (denoted as Conv1 – Conv5, respectively) and three fully connected layers (denoted as FC1 – FC3). The activation of each image was extracted and vectorized from each layer using the DNNBrain toolbox[86]. The visual similarity matrix was generated for each layer with the value cell $(i,j)$ denoting dissimilarity (1 – Pearson's $r$) between the corresponding activation vectors of object $i$ and object $j$.

**Regression analysis**

To examine the relationship between object representation and manipulation properties, we vectorized the lower triangular matrices of the subjective dissimilarity matrix and corresponding manipulation and visual dissimilarity matrices. We used the subjective dissimilarity vector as the dependent variable and corresponding z-transformed manipulation dissimilarity vectors as independent variables. The resultant Coefficient of Determination ($R^2$) was calculated as the measure of goodness of fit, reflecting the proportion of variance in the subjective dissimilarity matrix explained by corresponding manipulation dissimilarity matrices. Additionally, by including more than one group of IVs in the regression, we calculated the proportion of variance collectively explained by these groups. Further, by including various groups of IVs separately and comparing the resultant $R^2$, we calculated their respective unique and shared contributions.

Specifically, for each type of objects (the everyday and the novel objects), with the subjective dissimilarity vector as the dependent variable, we examined four (sets of) regression analyses. The first regression included all the dissimilarity vectors of manipulation features as IVs. The $R^2$ of this regression illustrated the association between manipulation similarity and subjective similarity (a + b + d + e + f + g in Figure 5). The second set of regressions included either affordance or situatedness features only as IVs, exploring the association between each type of manipulation features (a + d + e + g or b + d + f + g in Figure 5) and subjective similarity. The difference between the $R^2$ of the first regression and that of the corresponding regression in the second set indexed the amount of variation uniquely explained by the other category (b + f or a + e in Figure 5). The difference between the $R^2$ of the first regression and the sum of the uniquely explained variance of either category of manipulation features indexed the amount of variance explained by both categories of manipulation features (g + d in Figure 5). The third and the fourth set of regressions took visual features into account. The third regression was a hierarchical multiple regression analysis with the visual dissimilarity vectors entered in the first step and the manipulation dissimilarity vectors entered in the second step. The visual vectors were vectorized visual dissimilarity matrices of each layer of the AlexNet. The change in $R^2$ indicated the amount of variance uniquely explained by manipulation similarity (a + b + d in Figure 5), and the difference in the $R^2$ between this regression and the first one indicated the variance uniquely explained by visual similarity (c in Figure 5). The difference between the $R^2$ of this regression and the sum of variances uniquely explained by either visual (c in Figure 5) or manipulation similarity (a + b + d in Figure 5) indicated the variance explained by both visual and manipulation similarity (e + f + g in Figure 5). A fourth set of regression consisted of two hierarchical multiple regression analyses with the visual similarity vectors in the first step and one category of manipulation similarity vectors in the second step. The change of $R^2$ after introducing the second step in either regression in this set indicated the variance uniquely explained by the corresponding category of manipulation features after taking visual features into account (a + d or b + d in Figure 5), and the difference between them and the variances uniquely explained by manipulation similarity (a + b + d in Figure 5) indicated the variance explained uniquely by that category of embodied features (a or b in Figure 5), which further enabled the calculation of the variance explained by both categories of embodied features but not the visual features (d in Figure 5).

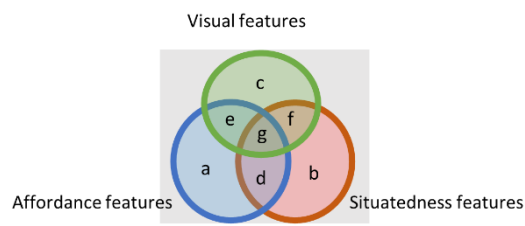

**Figure 5. Illustration of the estimation of variance explained by visual, affordance, and situatedness features.**

All the statistical tests reported in the present study were two-sided.

**Comparison between representational geometry**

To compare representational geometries based on subjective similarity and embodied and visual features, we needed to map the subjective and feature-based representational geometries into spaces of the same dimensionality. For this purpose, we estimated the dimensionality of the subjective object space based on subjective similarity, then extracted principal components from the embodied and visual features by principal component analysis (PCA), constructed hypothetical spaces of the same dimensionality by combining these principal components, and examined how well the representational geometry in the constructed spaces drawing on various combinations of embodied and visual components matched with that in the subjective object space. PCA was conducted with the *pca* function in Matlab R2020a.

### Construction of the subjective object space

We estimated the dimensionality of the subjective object space by submitting the group subjective dissimilarity matrix to non-metric multidimensional scaling (MDS)[41–44], which maps the data points from a high-dimensional space to lower-dimensional spaces by approximating a monotonic transformation of the pairwise differences to preserve the relative distance in the high dimensional space between object pairs in the target low dimensional space. Without a priori hypothesis, we determined the dimensionality of the subjective object space based on the MDS stress curve. MDS stress is a quantitative measure of the dissimilarity in the high dimensional space and the corresponding distances of the resultant low dimensional space. It is defined as the residual variance of the monotone regression of distance upon dissimilarity, normalized by the sum of squares of the inter-point distances. The smaller the stress, the better the fit between the two spaces. Our MDS stress curve reached the rule of thumb goodness-of-fit criterion (stress < 0.05)[45] before seven dimensions, which means the coffee mugs were satisfactorily represented in the first seven dimensions. Therefore, in the following analysis, we took seven as the estimated dimensionality of the subjective object space, with the coordinates of each mug in the first seven MDS dimensions as their coordinates in the subjective object space. Non-metric MDS analysis was conducted with *midscale* function in Matlab R2020a (The MathWorks Inc).

### Construction of hypothetical object spaces

We then tried to construct hypothetical object spaces of the same dimensionality as the subjective object space. We first conducted PCA on the group average embodied atom scores and AlexNet activation, separately. To decide which component to retain in the search pool, we calculated their corresponding effective dimensionality[46,47] following formula (1). The EDs were 5.60 and 18.64, respectively, consistent with the impression from the scree plot (Supplementary Figure S3), indicating the first 6 embodied components and 19 visual components derived from the PCA effectively reflected variance between objects in the embodied and visual feature spaces, respectively. Therefore, we included these 25 principal components in the search pool and constructed hypothetical object spaces of varied dimensionality by drawing the same number of components from it.

$$ED = \frac{1}{(explained\_d' \times explained\_d)} \qquad (1)$$

where $explained\_d$ is the vector of normalized explained variance of each component in the PCA analysis.

*Searching for best-fit between the subjective and the alternative spaces*

We quantified the similarity between the subjective object space and hypothetical spaces using Procrustes distances[48]. In Procrustes analysis, the distance describes the shape differences between two point sets, A and B, by transforming A into C to match B via scale changes, rotations, and translations. The Procrustes distance is the sum of squared differences between C and B. The smaller the Procrustes distances, the more similar the representational geometries of the same set of objects in the two spaces in question, and thus the more similar the two spaces. To achieve the minimal Procrustes distance, we exhausted all possible combinations of the embodied and visual components from the search pool with a total number equal to or smaller than seven (to either match the dimensionality of subjective object spaces or one of its sub-spaces). We searched for the hypothetical object space with the minimum Procrustes distance to the subjective object space, and considered it an indication of necessity in representing subjective object space if embodied components were presented in the best-match constructed object spaces. Procrustes analysis was conducted with *procrustes* function in Matlab R2020a.

**2D visualizations of the representational geometry in the subjective and other spaces**

To visualize the representational geometries in different object spaces, 2D visualizations were generated for the subjective object space, the embodied feature space, and the visual feature space. The visualization for the subjective object space was derived from the 2D non-metric MDS of subjective similarity data; for the embodied and visual feature spaces, the manipulation judgments or the AlexNet activations were first submitted to PCA, and the resultant component scores were then subjected to 2D t-SNE projection. The mug images were framed in the same color scheme as in the dendrogram in Figure 1b, which reflects the hierarchical clustering based on the subjective similarity data.

## Acknowledgements

We thank Xiaoyue Chi and Ziqian Xu for their assistance in data collection. Shan Xu discloses support for the research of this work from Natural Science Foundation of China (32371099), Xinran Feng discloses support for the research of this work from Shuimu Scholar Program of Tsinghua University and China Postdoctoral International Exchange Program (YJ20220273), and Jia Liu discloses support for the research of this work from Beijing Municipal Science & Technology Commission, Administrative Commission of Zhongguancun Science Park (Z221100002722012), and Double First-Class Initiative Funds for Discipline Construction.

## Data availability

The data that support the findings of this study are available at https://osf.io/5e9r7/.

## Code availability

The codes that support the findings of this study are available at https://osf.io/5e9r7/.

## Author contribution statement

S. X. contributed to the conceptualization, investigation, formal analysis, data curation, writing-original draft, writing – review and editing, visualization, and fund acquisition. X. F. contributed to the conceptualization, investigation, writing-original draft, writing–review and editing, visualization, and fund acquisition. Y. L. contributed to the investigation, formal analysis, data curation, writing – review and editing. J. L. contributed to the conceptualization, writing–review and editing, supervision, and fund acquisition. S. X. and X. F. contributed equally to the study.

Correspondence should be addressed to J. L. (liujiathu@tsinghua.edu.cn).

**Competing interests**

**Reference**

1. DiCarlo, J. J., Zoccolan, D. & Rust, N. C. How Does the Brain Solve Visual Object Recognition? *Neuron* **73**, 415–434 (2012).
2. DiCarlo, J. J. & Cox, D. D. Untangling invariant object recognition. *Trends Cogn. Sci.* **11**, 333–341 (2007).
3. Huang, T., Song, Y. & Liu, J. Real-world size of objects serves as an axis of object space. *Commun. Biol.* **5**, 749 (2022).
4. Kriegeskorte, N. & Kievit, R. A. Representational geometry: integrating cognition, computation, and the brain. *Trends Cogn. Sci.* **17**, 401–412 (2013).
5. Bao, P., She, L., McGill, M. & Tsao, D. Y. A map of object space in primate inferotemporal cortex. *Nature* **583**, 103–108 (2020).
6. Blumenthal, A., Stojanoski, B., Martin, C. B., Cusack, R. & Köhler, S. Animacy and real-world size shape object representations in the human medial temporal lobes. *Hum. Brain Mapp.* **39**, 3779–3792 (2018).
7. Grill-Spector, K. & Weiner, K. S. The functional architecture of the ventral temporal cortex and its role in categorization. *Nat. Rev. Neurosci.* **15**, 536–548 (2014).
8. Hebart, M. N., Zheng, C. Y., Pereira, F. & Baker, C. I. Revealing the multidimensional mental representations of natural objects underlying human similarity judgements. *Nat. Hum. Behav.* **4**, 1173–1185 (2020).
9. Julian, J. B., Ryan, J. & Epstein, R. A. Coding of Object Size and Object Category in Human

Visual Cortex. *Cereb. Cortex* bhw150 (2016) doi:10.1093/cercor/bhw150.

10. Konkle, T. & Oliva, A. A Real-World Size Organization of Object Responses in Occipitotemporal Cortex. *Neuron* **74**, 1114–1124 (2012).

11. Sha, L. *et al.* The Animacy Continuum in the Human Ventral Vision Pathway. *J. Cogn. Neurosci.* **27**, 665–678 (2015).

12. Yargholi, E. & Op De Beeck, H. Category Trumps Shape as an Organizational Principle of Object Space in the Human Occipitotemporal Cortex. *J. Neurosci.* **43**, 2960–2972 (2023).

13. Brunswik, E. *Perception and the Representative Design of Psychological Experiments*. (University of California Press, Berkeley, 1956).

14. Bracci, S. & Op De Beeck, H. P. Understanding Human Object Vision: A Picture Is Worth a Thousand Representations. *Annu. Rev. Psychol.* **74**, 113–135 (2023).

15. Feng, X., Xu, S., Li, Y. & Liu, J. Body size as a metric for the affordable world. *eLife* **12**, RP90583 (2024).

16. Holler, D. E., Fabbri, S. & Snow, J. C. Object responses are highly malleable, rather than invariant, with changes in object appearance. *Sci. Rep.* **10**, 4654 (2020).

17. Chen, J., Snow, J. C., Culham, J. C. & Goodale, M. A. What Role Does "Elongation" Play in "Tool-Specific" Activation and Connectivity in the Dorsal and Ventral Visual Streams? *Cereb. Cortex* **28**, 1117–1131 (2018).

18. Pellicano, A., Iani, C., Borghi, A. M., Rubichi, S. & Nicoletti, R. Simon-Like and Functional Affordance Effects with Tools: The Effects of Object Perceptual Discrimination and Object Action State. *Q. J. Exp. Psychol.* **63**, 2190–2201 (2010).

19. Phillips, J. C. & Ward, R. S-R correspondence effects of irrelevant visual affordance: Time course and specificity of response activation. *Vis. Cogn.* **9**, 540–558 (2002).

20. Tucker, M. & Ellis, R. On the Relations Between Seen Objects and Components of Potential Actions. *J. Exp. Psychol. Hum. Percept. Perform.* **24**, 830–846 (1998).

21. Adamson, R. Functional fixedness as related to problem solving - a repetition of 3 experiments. *J. Exp. Psychol.* **44**, 288–291 (1952).

22. Jeannerod, M. Specialized channels for cognitive responses. *Cognition* **10**, 135–137 (1981).

23. Borghi, A. M. & Riggio, L. Stable and variable affordances are both automatic and flexible. *Front. Hum. Neurosci.* **9**, (2015).

24. Wilson, M. Six views of embodied cognition. *Psychon. Bull. Rev.* **9**, 625–636 (2002).

25. Gibson, J. J. *The Ecological Approach to Visual Perception: Classic Edition*. (Psychology press, 1979).

26. Monaco, S., Sedda, A., Cavina-Pratesi, C. & Culham, J. C. Neural correlates of object size and object location during grasping actions. *Eur. J. Neurosci.* **41**, 454–465 (2015).

27. Borghi, A. M. & Riggio, L. Sentence comprehension and simulation of object temporary, canonical and stable affordances. *Brain Res.* **1253**, 117–128 (2009).

28. Edelman, S. Representation is representation of similarities. *Behav. Brain Sci.* **21**, 449–467 (1998).

29. Bullock, I. M., Ma, R. R. & Dollar, A. M. A Hand-Centric Classification of Human and Robot Dexterous Manipulation. *IEEE Trans. Haptics* **6**, 129–144 (2013).

30. Feix, T., Romero, J., Schmiedmayer, H.-B., Dollar, A. M. & Kragic, D. The GRASP Taxonomy of Human Grasp Types. *IEEE Trans. Hum.-Mach. Syst.* **46**, 66–77 (2016).

31. Santello, M., Flanders, M. & Soechting, J. F. Postural Hand Synergies for Tool Use. *J. Neurosci.* **18**, 10105–10115 (1998).

32. Matheson, H. E., Garcea, F. E. & Buxbaum, L. J. Scene context shapes category representational geometry during processing of tools. *Cortex* **141**, 1–15 (2021).

33. Rowe, P. J., Haenschel, C., Kosilo, M. & Yarrow, K. Objects rapidly prime the motor system when located near the dominant hand. *Brain Cogn.* **113**, 102–108 (2017).

34. Krizhevsky, A. One weird trick for parallelizing convolutional neural networks. Preprint at http://arxiv.org/abs/1404.5997 (2014).

35. Krizhevsky, A., Sutskever, I. & Hinton, G. E. ImageNet classification with deep convolutional neural networks. *Commun. ACM* **60**, 84–90 (2017).

36. Guclu, U. & Van Gerven, M. A. J. Deep Neural Networks Reveal a Gradient in the Complexity of Neural Representations across the Ventral Stream. *J. Neurosci.* **35**, 10005–10014 (2015).

37. Huang, T., Zhen, Z. & Liu, J. Semantic Relatedness Emerges in Deep Convolutional Neural Networks Designed for Object Recognition. *Front. Comput. Neurosci.* **15**, 625804 (2021).

38. Liu, X., Zhen, Z. & Liu, J. Hierarchical Sparse Coding of Objects in Deep Convolutional Neural Networks. *Front. Comput. Neurosci.* **14**, 578158 (2020).

39. Wen, H. *et al.* Neural Encoding and Decoding with Deep Learning for Dynamic Natural Vision.

*Cereb. Cortex* **28**, 4136–4160 (2018).

40. Xu, S., Zhang, Y., Zhen, Z. & Liu, J. The Face Module Emerged in a Deep Convolutional Neural Network Selectively Deprived of Face Experience. *Front. Comput. Neurosci.* **15**, 626259 (2021).

41. Kruskal, J. B. Nonmetric multidimensional scaling: A numerical method. *Psychometrika* **29**, 115–129 (1964).

42. Kruskal, J. B. Multidimensional scaling by optimizing goodness of fit to a nonmetric hypothesis. *Psychometrika* **29**, 1–27 (1964).

43. Shepard, R. N. The analysis of proximities: Multidimensional scaling with an unknown distance function. I. *Psychometrika* **27**, 125–140 (1962).

44. Shepard, R. N. The analysis of proximities: Multidimensional scaling with an unknown distance function. II. *Psychometrika* **27**, 219–246 (1962).

45. Hair, J. F. *Multivariate Data Analysis*. (2009).

46. Farrell, M., Recanatesi, S., Moore, T., Lajoie, G. & Shea-Brown, E. Gradient-based learning drives robust representations in recurrent neural networks by balancing compression and expansion. *Nat. Mach. Intell.* **4**, 564–573 (2022).

47. Rajan, K., Abbott, L. F. & Sompolinsky, H. Stimulus-dependent suppression of chaos in recurrent neural networks. *Phys. Rev. E* **82**, 011903 (2010).

48. Kendall, D. G. A Survey of the Statistical Theory of Shape. *Stat. Sci.* **4**, 87–99 (1989).

49. Nitzken, M. J. *et al.* Shape Analysis of the Human Brain: A Brief Survey. *IEEE J. Biomed. Health Inform.* **18**, 1337–1354 (2014).

50. Zaidi, K. F. & Harris-Love, M. Upper extremity kinematics: development of a quantitative measure of impairment severity and dissimilarity after stroke. *PeerJ* **11**, e16374 (2023).

51. Cavdan, M., Goktepe, N., Drewing, K. & Doerschner, K. Assessing the representational structure of softness activated by words. *Sci. Rep.* **13**, 8974 (2023).

52. Struch, N., Schwartz, S. H. & Van Der Kloot, W. A. Meanings of Basic Values for Women and Men: A Cross-Cultural Analysis. *Pers. Soc. Psychol. Bull.* **28**, 16–28 (2002).

53. Tenenbaum, J. B., Silva, V. D. & Langford, J. C. A Global Geometric Framework for Nonlinear Dimensionality Reduction. *Science* **290**, 2319–2323 (2000).

54. Vandenberg, S. & Kuse, A. Mental rotations, a group test of 3-dimensional spatial visualization.

*Percept. Mot. Skills* **47**, 599–604 (1978).

55. Almeida, J., Mahon, B. Z. & Caramazza, A. The Role of the Dorsal Visual Processing Stream in Tool Identification. *Psychol. Sci.* **21**, 772–778 (2010).

56. Bergström, F., Wurm, M., Valério, D., Lingnau, A. & Almeida, J. Decoding stimuli (tool-hand) and viewpoint invariant grasp-type information. *Cortex* **139**, 152–165 (2021).

57. Buchwald, M., Przybylski, Ł. & Króliczak, G. Decoding Brain States for Planning Functional Grasps of Tools: A Functional Magnetic Resonance Imaging Multivoxel Pattern Analysis Study. *J. Int. Neuropsychol. Soc.* **24**, 1013–1025 (2018).

58. Chao, L. L. & Martin, A. Representation of Manipulable Man-Made Objects in the Dorsal Stream. *NeuroImage* **12**, 478–484 (2000).

59. Errante, A., Ziccarelli, S., Mingolla, G. P. & Fogassi, L. Decoding grip type and action goal during the observation of reaching-grasping actions: A multivariate fMRI study. *NeuroImage* **243**, 118511 (2021).

60. Gerlach, C., Law, I. & Paulson, O. B. When Action Turns into Words. Activation of Motor-Based Knowledge during Categorization of Manipulable Objects. *J. Cogn. Neurosci.* **14**, 1230–1239 (2002).

61. Grèzes, J. & Decety, J. Does visual perception of object afford action? Evidence from a neuroimaging study. *Neuropsychologia* **40**, 212–222 (2002).

62. He, H., Zhuo, Y., He, S. & Zhang, J. The transition from invariant to action-dependent visual object representation in human dorsal pathway. *Cereb. Cortex* **32**, 5503–5511 (2022).

63. Matić, K., Op De Beeck, H. & Bracci, S. It's not all about looks: The role of object shape in parietal representations of manual tools. *Cortex* **133**, 358–370 (2020).

64. Rizzolatti, G. *et al.* Functional-organization of inferior area-6 in the macaque monkey .2. area f5 and the control of distal movements. *Exp. Brain Res.* **71**, 491–507 (1988).

65. Valyear, K. F., Gallivan, J. P., McLean, D. A. & Culham, J. C. fMRI Repetition Suppression for Familiar But Not Arbitrary Actions with Tools. *J. Neurosci.* **32**, 4247–4259 (2012).

66. Beer, R. D. The Dynamics of Active Categorical Perception in an Evolved Model Agent. *Adapt. Behav.* **11**, 209–243 (2003).

67. Xu, S., Liu, X., Almeida, J. & Heinke, D. The contributions of the ventral and the dorsal visual streams to the automatic processing of action relations of familiar and unfamiliar object pairs.

*NeuroImage* **245**, 118629 (2021).

68. Kristensen, S., Garcea, F. E., Mahon, B. Z. & Almeida, J. Temporal Frequency Tuning Reveals Interactions between the Dorsal and Ventral Visual Streams. *J. Cogn. Neurosci.* **28**, 1295–1302 (2016).

69. Mahon, B. Z., Kumar, N. & Almeida, J. Spatial Frequency Tuning Reveals Interactions between the Dorsal and Ventral Visual Systems. *J. Cogn. Neurosci.* **25**, 862–871 (2013).

70. Milner, A. D. How do the two visual streams interact with each other? *Exp. Brain Res.* **235**, 1297–1308 (2017).

71. Cohen, N. R., Cross, E. S., Tunik, E., Grafton, S. T. & Culham, J. C. Ventral and dorsal stream contributions to the online control of immediate and delayed grasping: A TMS approach. *Neuropsychologia* **47**, 1553–1562 (2009).

72. Milner, A. D. & Goodale, M. A. Two visual systems re-viewed. *Neuropsychologia* **46**, 774–785 (2008).

73. Jiang, L., Zhu, Y. & Liu, J. Information is asymmetry: spatial relations were encoded by asymmetric mnemonic manifolds. Preprint at https://doi.org/10.1101/2024.03.13.584855 (2024).

74. Ma, H., Jiang, L., Liu, T. & Liu, J. From sensory to perceptual manifold: the twist of neural geometry. Preprint at https://doi.org/10.1101/2023.10.02.559721 (2023).

75. Vaswani, A. *et al.* Attention Is All You Need. Preprint at http://arxiv.org/abs/1706.03762 (2023).

76. Radford, A., Narasimhan, K., Salimans, T. & Sutskever, I. Improving language understanding by generative pre-training. (2018).

77. Mikolov, T., Chen, K., Corrado, G. & Dean, J. Efficient Estimation of Word Representations in Vector Space. Preprint at http://arxiv.org/abs/1301.3781 (2013).

78. Wang, Y., Hou, Y., Che, W. & Liu, T. From static to dynamic word representations: a survey. *Int. J. Mach. Learn. Cybern.* **11**, 1611–1630 (2020).

79. Gupta, A., Savarese, S., Ganguli, S. & Fei-Fei, L. Embodied intelligence via learning and evolution. *Nat. Commun.* **12**, 5721 (2021).

80. Duan, J., Yu, S., Tan, H. L., Zhu, H. & Tan, C. A Survey of Embodied AI: From Simulators to Research Tasks. Preprint at http://arxiv.org/abs/2103.04918 (2022).

81. Deng, J. *et al.* ImageNet: A large-scale hierarchical image database. in *2009 IEEE Conference*

*on Computer Vision and Pattern Recognition* 248–255 (IEEE, Miami, FL, 2009). doi:10.1109/CVPR.2009.5206848.

82. Peirce, J. *et al.* PsychoPy2: Experiments in behavior made easy. *Behav. Res. Methods* **51**, 195–203 (2019).

83. Oldfield, R. C. The assessment and analysis of handedness: the Edinburgh inventory. *Neuropsychologia* **9**, 97–113 (1971).

84. Yang, N., Waddington, G., Adams, R. & Han, J. Translation, cultural adaption, and test–retest reliability of Chinese versions of the Edinburgh Handedness Inventory and Waterloo Footedness Questionnaire. *Laterality Asymmetries Body Brain Cogn.* **23**, 255–273 (2018).

85. Krizhevsky, A., Sutskever, I. & Hinton, G. E. Imagenet classification with deep convolutional neural networks. *Adv. Neural Inf. Process. Syst.* **25**, (2012).

86. Chen, X. *et al.* DNNBrain: a unifying toolbox for mapping deep neural networks and brains. *Front. Comput. Neurosci.* **14**, 580632 (2020).

**Supplementary material 1: Stimuli pool**

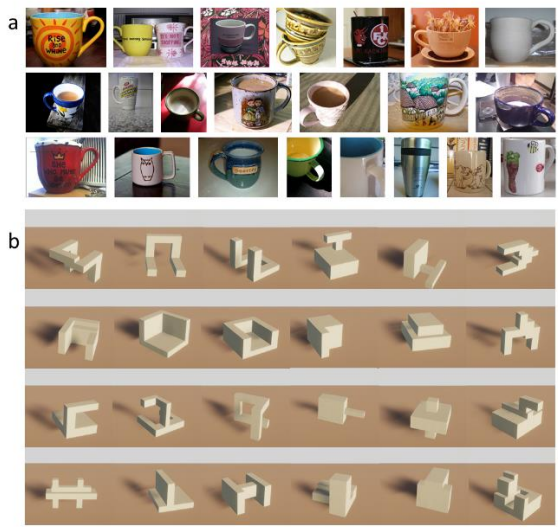

**Supplementary Figure S1.** *Stimuli pool.* **a**. The set of coffee mug images used as stimuli. **b**. The set of novel object images used as stimuli.

# Supplementary material 2: Action Atoms quantifying Embodied features

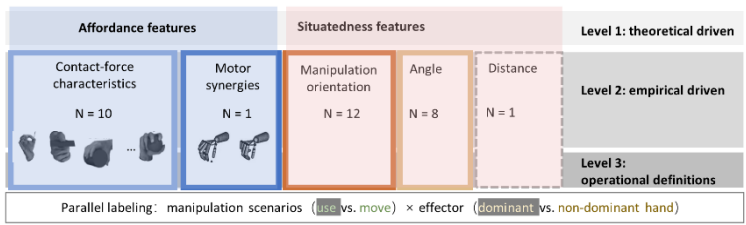

**Supplementary Figure S2.** *An overview of the action atoms*. A structured set of manipulation descriptions was developed to label the embodied features of objects. This set includes both affordance features which are largely dictated by the physical characteristics (e.g., shape and size) of the objects, and situatedness features which capture the variable embodied features specific to each manipulation instance, such as the agent-object spatial relation. Each category was then broken down into empirical-driven action atoms. Action atoms were further operationalized into judgment questions presented in the manipulation judgment task.

For coffee mug images, the questions were presented to the participants in a table for brevity. The questions were presented in simplified Chinese. Below is the English translation of the questions.

**Supplementary Table S1.** *List of questions used in manipulation judgment for coffee mug stimuli.*

| Scenario | Effector | Questions | | |
|---|---|---|---|---|
| To move | For the right hand (leave blank if not used) | Regarding hand-object contact | **Contact (1: Yes; 0: No)** | thumb |
| | | | | index finger |
| | | | | middle finger |
| | | | | ring finger |
| | | | | little finger |
| | | | | palm |
| | | | **Orientation of the index-thumb conjunction (1-12 o'clock), zero means perpendicular to the plan in question** | coronal |
| | | | | sagittal |
| | | | | horizontal |
| | | | **Orientation of the palm (1-12 o'clock), zero means perpendicular to the plan in question** | coronal |
| | | | | sagittal |
| | | | | horizontal |
| | | | **Manipulation posture** | thumb opposing another finger (0: NA; 1: opposing pad; 2: opposing back of another finger; 3: opposing the lateral side of another finger) |
| | | | | the finger that opposing thumb (2-5: index, middle, ring, and little fingers; blank: NA) |
| | | | | distance between opposing fingers (estimation in cm, 0: touch; blank: NA) |
| | | | | palm curving diameter (estimation in cm; 10000: flat) |
| | | | | index finger curving diameter (estimation in cm; 10000: flat) |
| | | | | middle finger curving diameter (estimation in cm; 10000: flat) |
| | | Movement after contact | **Hand movement direction (0-12 o'clock, zero means perpendicular to the plan in question)** | coronal |
| | | | | sagittal |
| | | | | horizontal |
| | | | **Rotation of hand relative to wrist (0: none; 1: clockwise; 2: counterclockwise)** | |
| | | | **Elbow (0: none; 1: bending; 2: stretching)** | |
| | | **Contacting or pressing with pad of the finger, against the object or other part of the hand (1/0)** | | thumb |
| | | | | index finger |
| | | | | middle finger |
| | | | | ring finger |
| | | | | little finger |
| | | **Contacting or pressing with the lateral side of the finger, against the object or other part of the hand (1/0)** | | thumb |
| | | | | index finger |
| | | | | middle finger |
| | | | | ring finger |
| | | | | little finger |
| | | **Contacting or pressing with the palm or hand back, against the object or other part of the hand (1/0)** | | palm |
| | | | | handback |
| | For the left hand (leave blank if not used) | Regarding hand-object contact | **Contact (1: Yes; 0: No)** | thumb |
| | | | | index finger |
| | | | | middle finger |
| | | | | ring finger |
| | | | | little finger |
| | | | | palm |

Supplementary Table S1: Continued.

| | | | | |
|---|---|---|---|---|
| | | | Orientation of the index-thumb conjunction (1-12 o'clock), zero means perpendicular to the plan in question | coronal |
| | | | | sagittal |
| | | | | horizontal |
| | | | Orientation of the palm (1-12 o'clock), zero means perpendicular to the plan in question | coronal |
| | | | | sagittal |
| | | | | horizontal |
| | | | Manipulation posture | thumb opposing another finger (0: NA; 1: opposing pad; 2: opposing back of another finger; 3: opposing the lateral side of another finger) |
| | | | | the finger that opposing thumb (2-5: index, middle, ring, and little fingers; blank: NA) |
| | | | | distance between opposing fingers (estimation in cm, 0: touch; blank: NA) |
| | | | | palm curving diameter (estimation in cm; 10000: flat) |
| | | | | index finger curving diameter (estimation in cm; 10000: flat) |
| | For the left hand (leave blank if not used) | | | middle finger curving diameter (estimation in cm; 10000: flat) |
| To move | | Movement after contact | Hand movement direction (0-12 o'clock, zero means perpendicular to the plan in question) | coronal |
| | | | | sagittal |
| | | | | horizontal |
| | | | Rotation of hand relative to wrist (0: none; 1: clockwise; 2: counterclockwise) | |
| | | | Elbow (0: none; 1: bending; 2: stretching) | |
| | | Contacting or pressing with pad of the finger, against the object or other part of the hand (1/0) | | thumb |
| | | | | index finger |
| | | | | middle finger |
| | | | | ring finger |
| | | | | little finger |
| | | Contacting or pressing with the lateral side of the finger, against the object or other part of the hand (1/0) | | thumb |
| | | | | index finger |
| | | | | middle finger |
| | | | | ring finger |
| | | | | little finger |
| | | Contacting or pressing with the palm or hand back, against the object or other part of the hand (1/0) | | palm |
| | | | | handback |
| To use | For the right hand (leave blank if not used) | Regarding hand-object contact | Contact (1: Yes; 0: No) | thumb |
| | | | | index finger |
| | | | | middle finger |
| | | | | ring finger |
| | | | | little finger |
| | | | | palm |
| | | | Orientation of the index-thumb conjunction (1-12 o'clock), zero means perpendicular to the plan in question | coronal |
| | | | | sagittal |
| | | | | horizontal |
| | | | Orientation of the palm (1-12 o'clock), zero means perpendicular to the plan in question | coronal |
| | | | | sagittal |
| | | | | horizontal |
| | | | Manipulation posture | thumb opposing another finger (0: NA; 1: opposing pad; 2: opposing back of another finger; 3: opposing the lateral side of another finger) |
| | | | | distance between opposing fingers (estimation in cm, 0: touch; blank: NA) |
| | | | | palm curving diameter (estimation in cm; 10000: flat) |

Supplementary Table S1: Continued.

| | | | |
|---|---|---|---|
| | **Regarding hand-object contact** | **Manipulation posture** | index finger curving diameter (estimation in cm; 10000: flat) |
| | | | middle finger curving diameter (estimation in cm; 10000: flat) |
| | **Movement after contact** | **Hand movement direction (0-12 o'clock, zero means perpendicular to the plan in question)** | coronal |
| | | | sagittal |
| | | | horizontal |
| | | **Rotation of hand relative to wrist (0: none; 1: clockwise; 2: counterclockwise)** | |
| **For the right hand (leave blank if not used)** | | **Elbow (0: none; 1: bending; 2: stretching)** | |
| | **Contacting or pressing with pad of the finger, against the object or other part of the hand (1/0)** | | thumb |
| | | | index finger |
| | | | middle finger |
| | | | ring finger |
| | | | little finger |
| | **Contacting or pressing with the lateral side of the finger, against the object or other part of the hand (1/0)** | | thumb |
| | | | index finger |
| | | | middle finger |
| | | | ring finger |
| | | | little finger |
| | **Contacting or pressing with the palm or hand back, against the object or other part of the hand (1/0)** | | palm |
| | | | handback |

**To use**

| | | | |
|---|---|---|---|
| | | **Contact (1: Yes; 0: No)** | thumb |
| | | | index finger |
| | | | middle finger |
| | | | ring finger |
| | | | little finger |
| | | | palm |
| | | **Orientation of the index-thumb conjunction (1-12 o'clock), zero means perpendicular to the plan in question** | coronal |
| | | | sagittal |
| | | | horizontal |
| | **Regarding hand-object contact** | **Orientation of the palm (1-12 o'clock), zero means perpendicular to the plan in question** | coronal |
| | | | sagittal |
| | | | horizontal |
| **For the left hand (leave blank if not used)** | | **Manipulation posture** | thumb opposing another finger (0: NA; 1: opposing pad; 2: opposing back of another finger; 3: opposing the lateral side of another finger) |
| | | | the finger that opposing thumb (2-5: index, middle, ring, and little fingers; blank: NA) |
| | | | distance between opposing fingers (estimation in cm, 0: touch; blank: NA) |
| | | | palm curving diameter (estimation in cm; 10000: flat) |
| | | | index finger curving diameter (estimation in cm; 10000: flat) |
| | | | middle finger curving diameter (estimation in cm; 10000: flat) |
| | **Movement after contact** | **Hand movement direction (0-12 o'clock, zero means perpendicular to the plan in question)** | coronal |
| | | | sagittal |
| | | | horizontal |
| | | **Rotation of hand relative to wrist (0: none; 1: clockwise; 2: counterclockwise)** | |
| | | **Elbow (0: none; 1: bending; 2: stretching)** | |

Supplementary Table S1: Continued.

| | | | |
|---|---|---|---|
| To use | For the left hand (leave blank if not used) | Contacting or pressing with pad of the finger, against the object or other part of the hand (1/0) | thumb |
| | | | index finger |
| | | | middle finger |
| | | | ring finger |
| | | | little finger |
| | | Contacting or pressing with the lateral side of the finger, against the object or other part of the hand (1/0) | thumb |
| | | | index finger |
| | | | middle finger |
| | | | ring finger |
| | | | little finger |
| | | Contacting or pressing with the palm or hand back, against the object or other part of the hand (1/0) | palm |
| | | | handback |

**Supplementary Table S2**. *The mapping between action atoms and sub-atoms*

| | | | | Original report | | | | |
|---|---|---|---|---|---|---|---|---|
| | | | Action atom | Thumb abduction or adduction and opposition type | Power or precision grasp | Virtual finger 1 | Virtual finger 2 and others | Additional specification |
| Affordance | Contact-and-force based | Prehensile | Power grasp (palm opposition) with thumb abducted | (the thumb opposed to another finger) OR (the thumb opposed to the back of another finger) | (the inner side of the index finger applied force on the object) OR (the palm applied force on the object) OR (the inner side of the middle finger applied force on the object) | (the inner side of the thumb applied force on the object) OR (the pad of the thumb applied force on the object) | (the inner side of the index finger applied force on the object) OR (the inner side of the middle finger applied force on the object) OR (the pad of the index finger applied force on the object) OR (the pad of the middle finger applied force on the object) | |
| | | | Power grasp (palm opposition) with thumb adducted | (the thumb did not oppose any other finger) OR (the thumb opposed to the lateral side of another finger) | (the inner side of the index finger applied force on the object) OR (the palm applied force on the object) OR (the inner side of the index finger applied force on the object) | | (the inner side of the index finger applied force on the object) OR (the inner side of the middle finger applied force on the object) OR (the inner side of the ring finger applied force on the object) OR (the pad of the index finger applied force on the object) OR (the pad of the middle finger applied force on the object) OR (the pad of the ring finger applied force on the object) | (the palm showed a curvature with an estimated diameter less than 15 cm) OR (the index finger showed a curvature with an estimated diameter less than 15 cm) OR (the middle finger showed a curvature with an estimated diameter less than 15 cm) |
| | | | Power grasp (pad opposition) with thumb abducted and the index and (or) middle fingers forming one or two virtual fingers | the thumb opposed to another finger | (the inner side of the index finger applied force on the object) OR (the inner side of the index finger applied force on the object) | (the pad of thumb applied force on the object) OR (the inner side of the thumb applied force on the object) | ((the inner side of the index finger applied force on the object) OR (the inner side of the index finger applied force on the object)) AND (the inner side of the ring finder did not apply force on the object) AND (the inner side of the little finger did not apply force on the object) AND (the pad of the ring finger did not apply force on the object) AND (the pad of the little finder did not apply force on the object) | (the index finger showed a curvature with an estimated diameter less than 15 cm) OR (the middle finger showed a curvature with an estimated diameter less than 15 cm) |

Supplementary Table S2. (Continued)

| Affordance | Contact-and-force based | Prehensile | | | | | |
|---|---|---|---|---|---|---|---|
| | | | Power grasp (pad opposition) with thumb abducted and the other fingers forming more than two virtual fingers | the thumb opposed to another finger | (the inner side of the index finger applied force on the object) OR (the inner side of the index finger applied force on the object) | (the pad of thumb applied force on the object) OR (the inner side of the thumb applied force on the object) | (the inner side of the index finger applied force on the object) AND (the inner side of the index finger applied force on the object) | |
| | | | Intermediate side grasp | the thumb did not oppose any other finger | | the pad of thumb applied force on the object | (the pad of the index finger applied force on the object) OR (the pad of the middle finger applied force on the object) OR (the inner side of the index finger applied force on the object) OR (the inner side of the middle finger applied force on the object) | (the palm showed a curvature with an estimated diameter less than 15 cm) OR (the index finger showed a curvature with an estimated diameter less than 15 cm) OR (the middle finger showed a curvature with an estimated diameter less than 15 cm) |
| | | | Precision grasp (pad opposition) with thumb abducted and the index and the middle fingers forming one or two virtual fingers | the thumb opposed to another finger | (the palm did not apply force on the object) AND (the inner side of the index finger did not apply force on the object) AND (the inner side of the middle finger did not apply force on the object) AND (the inner side o the ring finger did not apply force on the object) AND (the inner side of the little finger did not apply force on the object) | the pad of thumb applied force on the object | (((the thumb was opposed to the index finger) AND (the pad of the index finger applied force on the object) AND (the pad of the middle finder did not apply to the object)) OR (the thumb was opposed to the middle finger) AND (the pad of the middle finger applied force on the object) AND (the pad of the index finder did not apply to the object)) OR ((the pad of the index finger applied force on the object) AND (the pad of the middle finger applied force on the object))) AND (the pad of the ring finger did not apply force to the object) AND (the pad of the little finger did not apply force on the object) | (the index finger showed a curvature with an estimated diameter less than 15 cm) OR (the middle finger showed a curvature with an estimated diameter less than 15 cm) |

Supplementary Table S2. (Continued)

| | | | | | | | | |
|---|---|---|---|---|---|---|---|---|
| Affordance | Contact-and-force based | Prehensile | Precision grasp (pad opposition) with thumb abducted and the other fingers forming more than two virtual fingers | the thumb opposed to another finger | (the palm did not apply force on the object) AND (the inner side of the index finger did not apply force on the object) AND (the inner side of the middle finger did not apply force on the object) AND (the inner side o the ring finger did not apply force on the object) AND (the inner side of the little finger did not apply force on the object) | the pad of thumb applied force on the object | (the pad of the index finger applied force on the object) AND (the pad of the middle finger applied force on the object) AND (the pad of the ring finger applied force on the object) | (the index finger showed a curvature with an estimated diameter less than 15 cm) OR (the middle finger showed a curvature with an estimated diameter less than 15 cm) |
| | | Nonprehensile | Open hand nonprehensile manipulation with fingers | the thumb did not oppose any other finger | NA | | (the inner side of the index applied force on the object) OR (the inner side of the middle finger applied force on the object) OR (the inner side of the ring finger applied force on the object) OR (the pad of the index finger applied on force on the object) OR (the pad of the middle finger applied force on the object) OR (the pad of the ring finger applied force on the object) OR (the inner side of the thumb applied force on the object) OR (the pad of the thumb applied force on the object) | NOT (arm flexion) AND (NOT (prehensile)) |
| | | | Open hand nonprehensile manipulation with palm | the thumb did not oppose any other finger | NA | | the palm applied force on the object | NOT (arm flexion) AND (NOT (prehensile)) AND ((the palm showed a curvature with an estimated diameter more than 10 cm) OR (the index finger showed a curvature with an estimated diameter more than 10 cm) OR (the middle finger showed a curvature with an estimated diameter more than 10 cm)) |

Supplementary Table S2. (Continued)

| | | | | | | | |
|---|---|---|---|---|---|---|---|
| **Affordance** | Contact-and-force based | Nonprehensile | Open hand nonprehensile manipulation with handback | the thumb did not oppose any other finger | NA | the handback applied force on the object | NOT (arm flexion) AND (NOT (prehensile)) AND ((the palm showed a curvature with an estimated diameter more than 10 cm) OR (the index finger showed a curvature with an estimated diameter more than 10 cm) OR (the middle finger showed a curvature with an estimated diameter more than 10 cm)) |
| | Kinematic-based | | posture closeness | NA | | (the palm showed a curvature with an estimated diameter less than 5 cm) OR (the index finger showed a curvature with an estimated diameter less than 5 cm) OR (the middle finger showed a curvature with an estimated diameter less than 5 cm)) | |
| **Situated** | **Posture orientation** | **Palm Orientation** | | | **Specification** | | |
| | | | Inward | | The palm faced (between 8 and 10 o'clock on the sagittal plane) AND (between 5 and 7 o'clock on the horizontal plane) in contacting the object | | |
| | | | Outward | | The palm faced (between 2 and 4 o'clock on the sagittal plane) AND (between 11 and 1 o'clock on the horizontal plane) in contacting the object | | |
| | | | Leftward | | The palm faced (between 7 and 11 o'clock on the coronal plane) AND (between 7 and 11 o'clock on the horizontal plane) in contacting the object | | |
| | | | Rightward | | The palm faced (between 2 and 5 o'clock on the coronal plane) AND (between 2 and 5 o'clock on the horizontal plane) in contacting the object | | |
| | | | Upward | | The palm faced (between 11 and 1 o'clock on the coronal plane) AND (between 11 and 1 o'clock on the sagittal plane) in contacting the object | | |
| | | | Downward | | The palm faced (between 5 and 7 o'clock on the coronal plane) AND (between 5 and 7 o'clock on the sagittal plane) in contacting the object | | |
| | | **Finger Orientation** | Inward | | The lateral side of the index finger towards the thumb faced (between 8 and 10 o'clock on the sagittal plane) AND (between 5 and 7 o'clock on the horizontal plane) in contacting the object | | |
| | | | Outward | | The lateral side of the index finger towards the thumb faced (between 2 and 4 o'clock on the sagittal plane) AND (between 11 and 1 o'clock on the horizontal plane) in contacting the object | | |
| | | | Leftward | | The lateral side of the index finger towards the thumb faced (between 7 and 11 o'clock on the coronal plane) AND (between 7 and 11 o'clock on the horizontal plane) in contacting the object | | |

Supplementary Table S2. (Continued)

| | | | | |
|---|---|---|---|---|
| Situated | Posture orientation | | Rightward | The lateral side of the index finger towards the thumb faced (between 2 and 5 o'clock on the coronal plane) AND (between 2 and 5 o'clock on the horizontal plane) in contacting the object |
| | | | Upward | The lateral side of the index finger towards the thumb faced (between 11 and 1 o'clock on the coronal plane) AND (between 11 and 1 o'clock on the sagittal plane) in contacting the object |
| | | | Downward | The lateral side of the index finger towards the thumb faced (between 5 and 7 o'clock on the coronal plane) AND (between 5 and 7 o'clock on the sagittal plane) in contacting the object |
| | Movement direction | hand movement direction | Inward | The hand moved (between 8 and 10 o'clock on the sagittal plane) AND (between 5 and 7 o'clock on the horizontal plane) after contacting the object |
| | | | Outward | The hand moved (between 2 and 4 o'clock on the sagittal plane) AND (between 11 and 1 o'clock on the horizontal plane) after contacting the object |
| | | | Leftward | The hand moved (between 7 and 11 o'clock on the coronal plane) AND (between 7 and 11 o'clock on the horizontal plane) after contacting the object |
| | | | Rightward | The hand moved (between 2 and 5 o'clock on the coronal plane) AND (between 2 and 5 o'clock on the horizontal plane) after contacting the object |
| | | | Upward | The hand moved (between 11 and 1 o'clock on the coronal plane) AND (between 11 and 1 o'clock on the sagittal plane) after contacting the object |
| | | | Downward | The hand moved (between 5 and 7 o'clock on the coronal plane) AND (between 5 and 7 o'clock on the sagittal plane) after contacting the object |
| | | Arm movement | Wrist Twist | clockwise vs counterclockwise |
| | | | Elbow Flexion | bending vs stretching |
| | Perceived distance | | | estimated distance by centimeters between the agent and the object |

Supplementary Table S3: List of questions used in manipulation judgment for novel objects.

| Scenario | Effector | Categories | Questions | Choices |
|---|---|---|---|---|
| To move | right | **Orientation of the index-thumb conjunction** | Regarding the orientation of the index-thumb conjunction, when in contact with the objects, is it inward or outward? | inward / outward / neither |
| | | | Regarding the orientation of the index-thumb conjunction, when in contact with the objects, is it leftward or rightward? | leftward / rightward / neither |
| | | | Regarding the orientation of the index-thumb conjunction, when in contact with the objects, is it upward or downward? | upward / downward / neither |
| | | **Orientation of the palm** | Regarding the orientation of the hand palm, when in contact with the objects, is it inward or outward? | inward / outward / neither |
| | | | Regarding the orientation of the hand palm, when in contact with the objects, is it leftward or rightward? | leftward / rightward / neither |
| | | | Regarding the orientation of the hand palm, when in contact with the objects, is it upward or downward? | upward / downward / neither |
| | | **Hand movement direction** | After contacting the object, will the hand move outward or inward? | inward / outward / neither |
| | | | After contacting the object, will the hand move leftward or rightward? | leftward / rightward / neither |
| | | | After contacting the object, will the hand move upward or downward? | upward / downward / neither |
| | | **Elbow posture** | Does the elbow bend or stretch to contact the object? | bend / stretch / neither |
| | | **Manipulation posture** | Which description matches the hand posture of object manipulation most? | a. The object will not move within the hand in the absence of wrist and arm effort. The hand is in a gripping posture — the palm applies force to the object, with the thumb flexed inward, opposing the other fingers<br>b. The object will not move within the hand in the absence of wrist and arm effort. The hand is in a gripping posture — the palm applies force to the object, with the thumb not flexed inward (in a relaxed or upright position). |

Supplementary Table S3 (Continued).

| | | | | |
|---|---|---|---|---|
| **To move** | **right** | **Manipulation posture** | Which description matches the hand posture of object manipulation most? | c. The object will not move within the hand in the absence of wrist and arm effort. The hand is in a gripping posture — the pads or sides of the fingers apply force on the object, while the palm does not, with the thumb flexed inward, opposing the index or middle finger (the ring and the little fingers do not touch the object). |
| | | | | d. The object will not move within the hand in the absence of wrist and arm effort. The hand is in a gripping posture— the pads or sides of the fingers apply force on the object (the palm does not apply force), with the thumb flexed inward and opposing multiple fingers and the ring finger touching the object. |
| | | | | e. The sides of the fingers apply force to pinch the object (the sides of the thumb and index finger, or the sides of the index and middle fingers), and other parts may assist in stabilization. |
| | | | | f. The object can move flexibly within the hand without wrist and arm effort. The hand is in a pinching posture— the pads of the fingers apply force, with the thumb and index or middle finger pinching together (the ring and the little fingers do not touch the object) |
| | | | | g. The object can move flexibly within the hand without wrist and arm effort. The hand is in a pinching posture— the pads of the fingers apply force, with the thumb and multiple fingers pinching together and the ring finger touching the object. |
| | | | | h. The fingers apply force to the object, with other parts not touching the object in actions such as pushing with the fingers or hooking with the fingers. |
| | | | | i. The palm (and/or the base of the fingers) applies force to the object in actions such as pushing with the palm or supporting with the palm, with other parts not touching the object. |
| | | | | j. The back of the hand (and/or the base of the fingers) applies force to the object in actions such as pushing with the back of the hand, with other parts not touching the object. |
| | | | How much does the end of finger curve relative to the palm? | Not more than 45 degrees. |
| | | | | More than 45 degrees |
| | | | How much do the interphalangeal joints curve? | Not more than 45 degrees. |
| | | | | More than 45 degrees |
| | **left** | **Orientation of the index-thumb conjunction** | Regarding the orientation of the index-thumb conjunction, when in contact with the objects, is it inward or outward? | inward |
| | | | | outward |
| | | | | neither |
| | | | Regarding the orientation of the index-thumb conjunction, when in contact with the objects, is it leftward or rightward? | leftward |
| | | | | rightward |
| | | | | neither |

Supplementary Table S3 (Continued).

| | | | | |
|---|---|---|---|---|
| To move | left | **Orientation of the index-thumb conjunction** | Regarding the orientation of the index-thumb conjunction, when in contact with the objects, is it upward or downward? | upward |
| | | | | downward |
| | | | | neither |
| | | **Orientation of the palm** | Regarding the orientation of the hand palm, when in contact with the objects, is it inward or outward? | inward |
| | | | | outward |
| | | | | neither |
| | | | Regarding the orientation of the hand palm, when in contact with the objects, is it leftward or rightward? | leftward |
| | | | | rightward |
| | | | | neither |
| | | | Regarding the orientation of the hand palm, when in contact with the objects, is it upward or downward? | upward |
| | | | | downward |
| | | | | neither |
| | | **Hand movement direction** | After contacting the object, will the hand move outward or inward? | inward |
| | | | | outward |
| | | | | neither |
| | | | After contacting the object, will the hand move leftward or rightward? | leftward |
| | | | | rightward |
| | | | | neither |
| | | | After contacting the object, will the hand move upward or downward? | upward |
| | | | | downward |
| | | | | neither |
| | | **Elbow posture** | Does the elbow bend or stretch to contact the object? | bend |
| | | | | stretch |
| | | | | neither |
| | | **Manipulation posture** | Which description matches the hand posture of object manipulation most? | a. The object will not move within the hand in the absence of wrist and arm effort. The hand is in a gripping posture— the palm applies force to the object, with the thumb flexed inward, opposing the other fingers |
| | | | | b. The object will not move within the hand in the absence of wrist and arm effort. The hand is in a gripping posture— the palm applies force to the object, with the thumb not flexed inward (in a relaxed or upright position). |
| | | | | c. The object will not move within the hand in the absence of wrist and arm effort. The hand is in a gripping posture— the pads or sides of the fingers apply force on the object, while the palm does not, with the thumb flexed inward, opposing the index or middle finger (the ring and the little fingers do not touch the object). |
| | | | | d. The object will not move within the hand in the absence of wrist and arm effort. The hand is in a gripping posture— the pads or sides of the fingers apply force on the object (the palm does not apply force), with the thumb flexed inward and opposing multiple fingers and the ring finger touching the object. |

Supplementary Table S3 (Continued).

| | | | | |
|---|---|---|---|---|
| **To move** | **left** | **Manipulation posture** | Which description matches the hand posture of object manipulation most? | e. The sides of the fingers apply force to pinch the object (the sides of the thumb and index finger, or the sides of the index and middle fingers), and other parts may assist in stabilization. |
| | | | | f. The object can move flexibly within the hand without wrist and arm effort. The hand is in a pinching posture— the pads of the fingers apply force, with the thumb and index or middle finger pinching together (the ring and the little fingers do not touch the object) |
| | | | | g. The object can move flexibly within the hand without wrist and arm effort. The hand is in a pinching posture— the pads of the fingers apply force, with the thumb and multiple fingers pinching together and the ring finger touching the object. |
| | | | | h. The fingers apply force to the object, with other parts not touching the object in actions such as pushing with the fingers or hooking with the fingers. |
| | | | | i. The palm (and/or the base of the fingers) applies force to the object in actions such as pushing with the palm or supporting with the palm, with other parts not touching the object. |
| | | | | j. The back of the hand (and/or the base of the fingers) applies force to the object in actions such as pushing with the back of the hand, with other parts not touching the object. |
| | | | How much does the end of finger curve relative to the palm? | Not more than 45 degrees. |
| | | | | More than 45 degrees |
| | | | How much do the interphalangeal joints curve? | Not more than 45 degrees. |
| | | | | More than 45 degrees |
| **To use** | **right** | **Orientation of the index-thumb conjunction** | Regarding the orientation of the index-thumb conjunction, when in contact with the objects, is it inward or outward? | inward |
| | | | | outward |
| | | | | neither |
| | | | Regarding the orientation of the index-thumb conjunction, when in contact with the objects, is it leftward or rightward? | leftward |
| | | | | rightward |
| | | | | neither |
| | | | Regarding the orientation of the index-thumb conjunction, when in contact with the objects, is it upward or downward? | upward |
| | | | | downward |
| | | | | neither |
| | | **Orientation of the palm** | Regarding the orientation of the hand palm, when in contact with the objects, is it inward or outward? | inward |
| | | | | outward |
| | | | | neither |
| | | | Regarding the orientation of the hand palm, when in contact with the objects, is it leftward or rightward? | leftward |
| | | | | rightward |
| | | | | neither |
| | | | Regarding the orientation of the hand palm, when in contact with the objects, is it upward or downward? | upward |
| | | | | downward |
| | | | | neither |

Supplementary Table S3 (Continued).

| | | | |
|---|---|---|---|
| | **Hand movement direction** | After contacting the object, will the hand move outward or inward? | Inward |
| | | | outward |
| | | | neither |
| | | After contacting the object, will the hand move leftward or rightward? | leftward |
| | | | rightward |
| | | | neither |
| | | After contacting the object, will the hand move upward or downward? | upward |
| | | | downward |
| | | | neither |
| | **Elbow posture** | Does the elbow bend or stretch to contact the object? | bend |
| | | | stretch |
| | | | neither |
| **To use** right | **Manipulation posture** | Which description matches the hand posture of object manipulation most? | a. The object will not move within the hand in the absence of wrist and arm effort. The hand is in a gripping posture— the palm applies force to the object, with the thumb flexed inward, opposing the other fingers |
| | | | b. The object will not move within the hand in the absence of wrist and arm effort. The hand is in a gripping posture— the palm applies force to the object, with the thumb not flexed inward (in a relaxed or upright position). |
| | | | c. The object will not move within the hand in the absence of wrist and arm effort. The hand is in a gripping posture— the pads or sides of the fingers apply force on the object, while the palm does not, with the thumb flexed inward, opposing the index or middle finger (the ring and the little fingers do not touch the object). |
| | | | d. The object will not move within the hand in the absence of wrist and arm effort. The hand is in a gripping posture— the pads or sides of the fingers apply force on the object (the palm does not apply force), with the thumb flexed inward and opposing multiple fingers and the ring finger touching the object. |
| | | | e. The sides of the fingers apply force to pinch the object (the sides of the thumb and index finger, or the sides of the index and middle fingers), and other parts may assist in stabilization. |
| | | | f. The object can move flexibly within the hand without wrist and arm effort. The hand is in a pinching posture— the pads of the fingers apply force, with the thumb and index or middle finger pinching together (the ring and the little fingers do not touch the object) |
| | | | g. The object can move flexibly within the hand without wrist and arm effort. The hand is in a pinching posture— the pads of the fingers apply force, with the thumb and multiple fingers pinching together and the ring finger touching the object. |
| | | | h. The fingers apply force to the object, with other parts not touching the object in actions such as pushing with the fingers or hooking with the fingers. |

Supplementary Table S3 (Continued).

| | | | |
|---|---|---|---|
| To use | | Manipulation posture | Which description matches the hand posture of object manipulation most? | i. The palm (and/or the base of the fingers) applies force to the object in actions such as pushing with the palm or supporting with the palm, with other parts not touching the object.<br><br>j. The back of the hand (and/or the base of the fingers) applies force to the object in actions such as pushing with the back of the hand, with other parts not touching the object. |
| | | | How much does the end of finger curve relative to the palm? | Not more than 45 degrees.<br>More than 45 degrees |
| | | | How much do the interphalangeal joints curve? | Not more than 45 degrees.<br>More than 45 degrees |
| | left | Orientation of the index-thumb conjunction | Regarding the orientation of the index-thumb conjunction, when in contact with the objects, is it inward or outward? | inward<br>outward<br>neither |
| | | | Regarding the orientation of the index-thumb conjunction, when in contact with the objects, is it leftward or rightward? | leftward<br>rightward<br>neither |
| | | | Regarding the orientation of the index-thumb conjunction, when in contact with the objects, is it upward or downward? | upward<br>downward<br>neither |
| | | Orientation of the palm | Regarding the orientation of the hand palm, when in contact with the objects, is it inward or outward? | inward<br>outward<br>neither |
| | | | Regarding the orientation of the hand palm, when in contact with the objects, is it leftward or rightward? | leftward<br>rightward<br>neither |
| | | | Regarding the orientation of the hand palm, when in contact with the objects, is it upward or downward? | upward<br>downward<br>neither |
| | | Hand movement direction | After contacting the object, will the hand move outward or inward? | inward<br>outward<br>neither |
| | | | After contacting the object, will the hand move leftward or rightward? | leftward<br>rightward<br>neither |
| | | | After contacting the object, will the hand move upward or downward? | upward<br>downward<br>neither |
| | | Elbow posture | Does the elbow bend or stretch to contact the object? | bend<br>stretch<br>neither |

Supplementary Table S3 (Continued).

| | | | | |
|---|---|---|---|---|
| **To use** | **left** | **Manipulation posture** | Which description matches the hand posture of object manipulation most? | a. The object will not move within the hand in the absence of wrist and arm effort. The hand is in a gripping posture— the palm applies force to the object, with the thumb flexed inward, opposing the other fingers |
| | | | | b. The object will not move within the hand in the absence of wrist and arm effort. The hand is in a gripping posture— the palm applies force to the object, with the thumb not flexed inward (in a relaxed or upright position). |
| | | | | c. The object will not move within the hand in the absence of wrist and arm effort. The hand is in a gripping posture— the pads or sides of the fingers apply force on the object, while the palm does not, with the thumb flexed inward, opposing the index or middle finger (the ring and the little fingers do not touch the object). |
| | | | | d. The object will not move within the hand in the absence of wrist and arm effort. The hand is in a gripping posture— the pads or sides of the fingers apply force on the object (the palm does not apply force), with the thumb flexed inward and opposing multiple fingers and the ring finger touching the object. |
| | | | | e. The sides of the fingers apply force to pinch the object (the sides of the thumb and index finger, or the sides of the index and middle fingers), and other parts may assist in stabilization. |
| | | | | f. The object can move flexibly within the hand without wrist and arm effort. The hand is in a pinching posture— the pads of the fingers apply force, with the thumb and index or middle finger pinching together (the ring and the little fingers do not touch the object) |
| | | | | g. The object can move flexibly within the hand without wrist and arm effort. The hand is in a pinching posture— the pads of the fingers apply force, with the thumb and multiple fingers pinching together and the ring finger touching the object. |
| | | | | h. The fingers apply force to the object, with other parts not touching the object in actions such as pushing with the fingers or hooking with the fingers. |
| | | | | i. The palm (and/or the base of the fingers) applies force to the object in actions such as pushing with the palm or supporting with the palm, with other parts not touching the object. |
| | | | | j. The back of the hand (and/or the base of the fingers) applies force to the object in actions such as pushing with the back of the hand, with other parts not touching the object. |
| | | | How much does the end of finger curve relative to the palm? | Not more than 45 degrees. |
| | | | | More than 45 degrees |
| | | | How much do the interphalangeal joints curve? | Not more than 45 degrees. |
| | | | | More than 45 degrees |

## Supplementary materials 3: MDS results of varied object space dimensionality

The main text illustrated the necessity of embodied features in constructing object space to resemble the 7D subjective object space. This necessity was replicated with subjective object spaces of other dimensionalities.

For the 6D subjective object space (the object space embedded in the first six MDS dimensions of the subjective similarity data), the minimal Procrustes distance was achieved with a constructed object space solely based on the second embodied component in mirroring a 1D sub-space of the subjective object space (the third MDS dimension). Consistently, in mirroring the entire 6D subjective object space, two embodied components (the second and the third PCA components) and four visual components constructed the best match.

Similarly, with the 5D subjective object space, the minimal Procrustes distance was achieved with a constructed object space solely based on the second embodied component in mirroring a 1D sub-space of the subjective object space (consisting of the third MDS dimension), while in mirroring the entire 5D subjective object space, three embodied components (the second, the third, and the fifth PCA components) and two visual components constructed the best match.

We also tested the 8D subjective object space, whose best match was a constructed object space constructed using the first, the second, the third, and the fifth embodied components and four visual components.

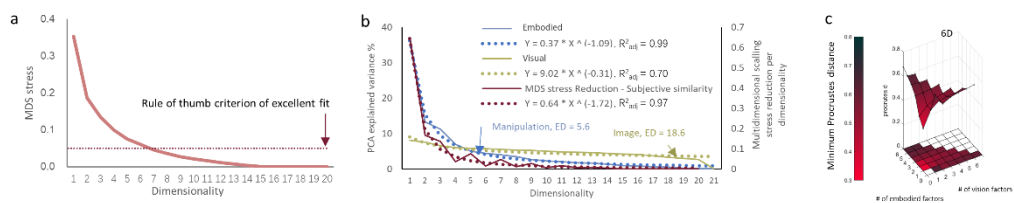

**Supplementary Figure S3.** *Dimensionality of spaces and MDS results of varied object space dimensionality*. **a.** Non-metric multidimensional scaling stress curve of the subjective similarity rating, indicating that the subjective object space can be satisfactorily scaled to a 7D space. The dashed line denotes the rule of thumb goodness-of-fit criterion (stress <0.05, Hair et al., 1998). The analyses in the main text therefore focused on the seven-dimensional subjective object space,

where each object was represented by 7D coordinates on the first 7 of all the possible MDS dimensions. **b.** The respective dimensionality of the embodied feature space and the visual feature space, and their comparison with the subjective object space. The PCA scree plots of the manipulation rating (the blue line) and the AlexNet feature extraction data (the yellow line) suggested higher dimensionalities of the visual feature space (ED = 18.6) than the embodied feature space (ED = 5.6). The curve of MDS stress reduction induced by introducing each subjective object space dimensionality (red line) took a similar shape to the manipulation curve. The dashed lines display the fitted power functions of the three curves, revealing that the distribution of explained variance across dimensionality in the manipulation curve and the subjective curve can both be nicely captured by power curves. In contrast, the fitted power function of the image space took a distinct shape, with smaller power exponent than the other two curves, suggesting a distinct high dimensionality and an even distribution of variance across dimensionality in the image space, different from the other two spaces. **c.** The best match, i.e., the minimum Procrustes distance (z-axis), between the subspaces of 6D subjective space and any equal-dimension feature space constructed by different numbers of manipulation (y-axis) and image components (x-axis). The lowest point in each surface denotes the overall best goodness-of-fit, and the x-y coordinates of each point indicate the number of embodied and physical components incorporated in the feature space. The minimal Procrustes distance was achieved with a constructed object space solely based on the second embodied component to mirror a 1D sub-space of the subjective object space (the third MDS dimension).